\documentclass[12pt]{article}
\usepackage{amsthm,amsmath,amssymb}
\usepackage{epsfig}
\usepackage{color}
\usepackage{multirow}
\usepackage{enumerate,verbatim}
\usepackage{bm}
\usepackage{setspace}
\usepackage{mathrsfs}
\usepackage{graphicx, subfig}
\usepackage{xcolor}
\usepackage{soul}
\usepackage{url}
\usepackage[normalem]{ulem} 
\allowdisplaybreaks[1]
\usepackage{cite}
\usepackage[round]{natbib}
\usepackage{setspace}
\usepackage{multirow,array}
\usepackage{natbib}
\usepackage{authblk}

\allowdisplaybreaks[1]

\renewcommand{\baselinestretch}{1.5}
\font\twelvesmc=cmcsc10 scaled\magstep1 
\newcommand{\smc}{\twelvesmc}

\newtheorem{lm}{ \underline{\smc Lemma}}

\newtheorem{thm}{\underline{\smc Theorem}}
\newtheorem{cl}{\underline{\smc Corollary}}
\newtheorem{pp}{\underline{\smc Proposition}}
\newtheorem*{remark}{\emph{\textmd{Remark}}}

\newtheorem{asmp}{Condition}
\newenvironment{cond}[1]
  {\renewcommand{\theasmp}{\ref{#1}$^*$}%
   \addtocounter{asmp}{-1}%
   \begin{asmp}}
  {\end{asmp}}

\newcommand{\bd}{\begin{document}}
\newcommand{\ed}{\end{document}}
\def\bse{\begin{eqnarray*}}
\def\ese{\end{eqnarray*}}
\def\be{\begin{eqnarray}}
\def\ee{\end{eqnarray}}
\newcommand{\ben}{\begin{eqnarray}}\newcommand{\een}{\end{eqnarray}}
\newcommand{\bn}{\begin{enumerate}}
\newcommand{\en}{\end{enumerate}}
\newcommand{\im}{\item}
\newcommand{\bc}{\begin{cases}}
\newcommand{\ec}{\end{cases}}
\newcommand{\bt}{\begin{tabular}}
\newcommand{\et}{\end{tabular}}
\newcommand{\bct}{\begin{center}}
\newcommand{\ect}{\end{center}}
\def\wt{\widetilde}
\def\diag{\hbox{diag}}
\def\wh{\widehat}
\def\AIC{\hbox{AIC}}
\def\BIC{\hbox{BIC}}
\newcommand{\tr}{{\rm tr}}
\newcommand{\cid}{\buildrel d \over \longrightarrow}
\newcommand{\cas}{\buildrel a.s \over \longrightarrow}
\newcommand{\cw}{\buildrel w \over \longrightarrow}
\newcommand{\cip}{\buildrel p \over \longrightarrow}
\newcommand{\cil}{\buildrel L^2 \over \longrightarrow}
\newcommand{\bl}{{\pmb l}}
\newcommand{\bs}{{\pmb s}}
\newcommand{\bh}{{\pmb h}}

\newcommand{\bda}{{\pmb a}}
\newcommand{\bdb}{{\pmb b}}
\newcommand{\bdc}{{\pmb c}}
\newcommand{\bdd}{{\pmb d}}
\newcommand{\bdf}{{\bf f}}
\newcommand{\bdg}{{\bf g}}
\newcommand{\bdx}{{\bf x}}
\newcommand{\bdu}{{\bf u}}
\newcommand{\bdv}{{\bf v}}

\newcommand{\bY}{{\bm Y}}
\newcommand{\bX}{{\bm X}}
\newcommand{\bZ}{{\bm Z}}

\newcommand{\red}[1]{\textcolor{red}{#1}}
\newcommand{\blue}[1]{\textcolor{blue}{#1}}
\newcommand{\green}[1]{\textcolor{teal}{#1}}
\newcommand{\brown}[1]{\textcolor{brown}{#1}}
\newcommand{\cyan}[1]{\textcolor{cyan}{#1}}
\newcommand{\ora}[1]{\textcolor{orange}{#1}}
\def\yongtao{\textcolor{blue}}

\newcommand{\Tau}{\overline{\mathrm{T}}}
\newcommand{\prev}{\mbox{\scriptscriptstyle{prev}}}

\newcommand{\bone}{{\pmb 1}}
\newcommand{\bdepsilon}{{\pmb\epsilon}}
\newcommand{\bdmu}{{\pmb \mu}}
\newcommand{\bdpsi}{{\pmb \psi}}
\newcommand{\bdalpha}{{\pmb \alpha}}
\newcommand{\bdbeta}{{\pmb \beta}}
\newcommand{\bdtheta}{{\pmb \theta}}
\newcommand{\bdTheta}{{\pmb \Theta}}
\newcommand{\bdGamma}{{\pmb \Gamma}}
\newcommand{\bdvartheta}{{\pmb \vartheta}}

\newcommand{\CA}{{\cal A}}
\newcommand{\CB}{{\cal B}}
\newcommand{\CC}{{\cal C}}
\newcommand{\CD}{{\cal D}}
\newcommand{\CE}{{\cal E}}
\newcommand{\CF}{{\cal F}}
\newcommand{\CG}{{\cal G}}
\newcommand{\CH}{{\cal H}}
\newcommand{\CJ}{{\cal J}}
\newcommand{\CK}{{\cal K}}
\newcommand{\CL}{{\cal L}}
\newcommand{\CM}{{\cal M}}
\newcommand{\CN}{{\cal N}}
\newcommand{\CO}{{\cal O}}
\newcommand{\CP}{{\cal P}}
\newcommand{\CQ}{{\cal Q}}
\newcommand{\CR}{{\cal R}}
\newcommand{\CS}{{\cal S}}
\newcommand{\CT}{{\cal T}}
\newcommand{\CU}{{\cal U}}
\newcommand{\CV}{{\cal V}}
\newcommand{\CW}{{\cal W}}
\newcommand{\CX}{{\cal X}}
\newcommand{\CY}{{\cal Y}}
\newcommand{\CZ}{{\cal Z}}
\newcommand{\BR}{{\bold R}}
\newcommand{\BA}{{\pmb A}}
\newcommand{\BB}{{\pmb B}}
\newcommand{\BC}{{\bm C}}
\newcommand{\BS}{{\pmb S}}
\newcommand{\BP}{{\mathbb P}}
\newcommand{\BM}{{\pmb M}}
\newcommand{\BG}{{\pmb G}}
\newcommand{\BT}{{\pmb T}}
\newcommand{\Bx}{{\pmb x}}
\newcommand{\BX}{{\pmb X}}
\newcommand{\BV}{{\pmb V}}
\newcommand{\BZ}{{\bold Z}}
\newcommand{\BY}{{\pmb Y}}
\newcommand{\FC}{\field{C}}
\newcommand{\la}{{\langle}}
\newcommand{\ra}{{\rangle}}
\newcommand{\E}{{\rm E}}
\newcommand{\bepsilon}{{\pmb varepsilon}}

\newcommand{\el}{L^2[0,1]}
\def\II{I\negthinspace I}
\def\III{I\negthinspace I\negthinspace I}
\renewcommand{\baselinestretch}{1.2}
\newcommand{\ep}{\vskip-.3cm\noindent\begin{flushright}
${{\hfill\llap{$\sqcup\!\!\!\!\sqcap$}}}$
\end{flushright}}
\long\def\symbolfootnote[#1]#2{\begingroup%
	\def\thefootnote{\fnsymbol{footnote}}\footnote[#1]{#2}\endgroup}

\def\wt{\widetilde}
\def\wh{\widehat}
\def\AIC{\hbox{AIC}}
\def\BIC{\hbox{BIC}}
\newcommand{\Appendix}
{
\def\thesection{Appendix~\Alph{section}}
\def\thesubsection{A.\arabic{subsection}}
}
\def\log{\hbox{log}}
\def\bias{\hbox{bias}}
\def\sd{\hbox{sd}}
\def\Siuu{\boldSigma_{i,uu}}
\def\ANNALS{{\it Annals of Statistics}}
\def\BIOK{{\it Biometrika}}
\def\whT{\widehat{\Theta}}
\def\STATMED{{\it Statistics in Medicine}}
\def\STATSCI{{\it Statistical Science}}
\def\JSPI{{\it Journal of Statistical Planning \& Inference}}
\def\JRSSB{{\it Journal of the Royal Statistical Society, Series B}}
\def\BMCS{{\it Biometrics}}
\def\COMMS{{\it Communications in Statistics, Theory \& Methods}}
\def\JQT{{\it Journal of Quality Technology}}
\def\STIM{{\it Statistics in Medicine}}
\def\TECH{{\it Technometrics}}
\def\AJE{{\it American Journal of Epidemiology}}
\def\JASA{{\it Journal of the American Statistical Association}}
\def\CDA{{\it Computational Statistics \& Data Analysis}}
\def\JCGS{{\it Journal of Computational and Graphical Statistics}}
\def\JCB{{\it Journal of Computational Biology}}
\def\BIOINF{{\it Bioinformatics}}
\def\JAMA{{\it Journal of the American Medical Association}}
\def\JNUTR{{\it Journal of Nutrition}}
\def\JCGS{{\it Journal of Computational and Graphical Statistics}}
\def\LETTERS{{\it Letters in Probability and Statistics}}
\def\JABES{{\it Journal of Agricultural, Biological and
                      Environmental Statistics}}
\def\JSPI{{\it Journal of Statistical Planning \& Inference}}
\def\BIOK{{\it Bio\-me\-tri\-ka}}
\def\JRSSB{{\it Journal of the Royal Statistical Society, Series B}}
\def\BMCS{{\it Biometrics}}
\def\COMMS{{\it Communications in Statistics, Series A}}
\def\JQT{{\it Journal of Quality Technology}}
\def\SCAN{{\it Scandinavian Journal of Statistics}}

\def\dfrac#1#2{{\displaystyle{#1\over#2}}}
\def\VS{{\vskip 3mm\noindent}}
\def\boxit#1{\vbox{\hrule\hbox{\vrule\kern6pt
          \vbox{\kern6pt#1\kern6pt}\kern6pt\vrule}\hrule}}
\def\refhg{\hangindent=20pt\hangafter=1}
\def\refmark{\par\vskip 2mm\noindent\refhg}
\def\naive{\hbox{naive}}
\def\itemitem{\par\indent \hangindent2\parindent \textindent}
\def\var{\hbox{var}}
\def\Var{\hbox{Var}}
\def\cov{\hbox{cov}}
\def\corr{\hbox{corr}}
\def\trace{\hbox{trace}}
\def\refhg{\hangindent=20pt\hangafter=1}
\def\refmark{\par\vskip 2mm\noindent\refhg}
\def\Normal{\hbox{Normal}}
\def\povr{\buildrel p\over\longrightarrow}
\def\ccdot{{\bullet}}
\def\bse{\begin{eqnarray*}}
\def\ese{\end{eqnarray*}}
\def\be{\begin{eqnarray}}
\def\ee{\end{eqnarray}}
\def\bq{\begin{equation}}
\def\eq{\end{equation}}
\def\bse{\begin{eqnarray*}}
\def\ese{\end{eqnarray*}}
\def\pr{\hbox{pr}}
\def\trans{^{\rm T}}
\def\myalpha{{\cal A}}
\def\th{^{th}}
\def\wi{{\hbox{\scriptsize WI}}}

\def\references{\bibliography{WeakSeparability_manuscript}}
\renewcommand\refname{ \noindent \bf References \vspace{0.0in}}
  \let\oldthebibliography=\thebibliography
  \let\endoldthebibliography=\endthebibliography
  \renewenvironment{thebibliography}[1]{
    \begin{oldthebibliography}{#1}
      \setlength{\parskip}{0ex}
      \setlength{\itemsep}{0ex}}
  {\end{oldthebibliography}}

\newcommand{\blind}{1}

\if1\blind
{
  \title{\bf Test of Weak Separability for Spatially Stationary Functional Field}
\author{Decai Liang$^1$,\ Hui Huang$^2$,\ Yongtao Guan$^3$\ \ and Fang Yao$^{4}$
\\

 {\noindent\em\small$^1$ School of Statistics and Data Science, Nankai University, China}
 
 {\noindent\em\small$^2$ School of Mathematics, Sun Yat-sen University, China}
 
 {\noindent\em\small$^3$ Department of Management Science, University of Miami, USA}
 
 {\noindent\em\small$^4$ School of Mathematical Sciences, Center for Statistical Science, Peking University, China}
 
}
}
 \fi

\if0\blind
{
  \title{\bf Test of Weak Separability for Spatially Stationary Functional Field}
\author{}
} \fi

\begin{document}
  \date{}
\maketitle

	%
\begin{abstract}		
\baselineskip 20pt

         For spatially dependent functional data, a generalized  Karhunen-Lo\`{e}ve expansion is commonly used to decompose data into an additive form of temporal components and spatially correlated coefficients. This structure provides a convenient model to investigate the space-time interactions, but may not hold for complex spatio-temporal processes. In this work, we introduce the concept of weak separability, and propose a formal test to examine its validity for non-replicated spatially stationary functional field. The asymptotic distribution of the test statistic that adapts to potentially diverging ranks is derived by constructing lag covariance estimation, which is easy to compute for practical implementation. We demonstrate the efficacy of the proposed test via simulations and illustrate its usefulness in two real examples: China PM$_{2.5}$ data and Harvard Forest data.  
\end{abstract}

	%
	%
	\par\vfill\noindent
	
	\par\medskip\noindent
	{\bf Short title}: Test of Weak Separability
	\newlength{\gnat}
	\setlength{\gnat}{20pt} \baselineskip=\gnat
	
	\section{Introduction}
Due to advances of modern sciences and technology, data with complex structures, especially those with varying temporal or/and spatial features, are more commonly collected and analyzed. Functional data analysis (FDA) has emerged as a prominent area and provides useful tools in a variety of applications, see \cite{RamsayJ.O.JamesO.2005Fda/} for an introduction, and \cite{Hsing2015Theoretical} for some theoretical foundations. 	
         In contrast to traditional multivariate analysis, functional data are viewed as realizations from infinite-dimensional stochastic processes or random functions, which require regularization, dimension reduction, and/or feature extraction pertaining to the modeling context. The Karhunen-Lo\`{e}ve (KL) expansion is one of the key means for such purposes. This leads to the much-celebrated functional principal component analysis (FPCA), see \cite{YaoFang2005FDAf}, \cite{hall2006properties}, \cite{LiYehua2010UCRF} and the references therein.
If the observed trajectories $X_i(t)$, $i=1,...,N$, are independently sampled from an $L^2$ process $X(t)\in L^2(\CT)$ with the mean function $\mu(t)=\E \{X_i(t)\}$ and covariance function $C(t_1,t_2)=\cov\{X_i(t_1), X_i(t_2)\}$, where $\CT$ is a compact interval in $\mathbb{R}$, then the KL expansion admits $X_i(t)= \mu(t) + \sum_{r=1}^\infty \xi_{ir}\psi_r(t)$, where $\psi_r(t)$ is the $r$th eigenfunction of $C(t_1,t_2)$ and $\xi_{ir}=\int_\CT \{X_i(t)-\mu(t)\}\psi_r(t)dt$ is the corresponding FPC score. 
In this model, the randomness of the process $X_i(t)$ is inherited by the FPC scores, i.e. projections of $X_i(t)$ onto the eigen-spaces, which are uncorrelated random variables whose  variances are the corresponding eigenvalues $\lambda_r$, where $\lambda_1>\lambda_2>\cdots \geq 0$ without loss of generality (w.l.o.g.). The FPCA provides an efficient method for low-rank approximation in the sense of capturing the most variability of the processes with the least components. In practice, one may often use a few leading FPCs for adequate approximation due to rapidly decaying eigenvalues. Note that the truncation serves as a tuning balance between model parsimony and fidelity to observed data, and could depend on the sample size as well as the complexity/smoothness of the underlying process.

In recent years, there has been emerging research on dependent functional data \citep[][among others]{hormann2010weakly, paul2011principal}, particularly on spatial functional data \citep[e.g.][]{paul2011principal, gromenko2012estimation, li2014functional, liu2017functional}, with applications including measurements of environmental factors (temperature, precipitation, air pollutants) across monitoring stations, vegetation index at different locations, and fMRI signals in biomedical imaging.
 To represent the curves observed at spatial locations $\bs_i, i=1,\ldots,N$, 
a widely used model in most of the above-mentioned work is the following generalized KL expansion:
\begin{equation}\label{spatial KL}
X(\bs_i,t)= \mu( \bs_i, t) + \sum_{r=1}^\infty \xi_{r}(\bs_i)\psi_r(t),
\end{equation}
where the FPC scores $\{\xi_{r}(\bs_i)_{i=1,\dots,N;r\geq 1}\}$ are assumed to be uncorrelated across the different components for any two locations $\bs_i$ and $\bs_{i'}$, i.e., $\cov \{ \xi_{j}(\bs_i),\xi_{k}(\bs_{i'}) \} =0$ for $j\not=k$. 
As a result, one assumes a simplified cross-covariance structure
\begin{equation}\label{spatial KL covariance}
C(\bs_1,t_1;\bs_2,t_2) = \sum_{r=1}^\infty \E\{ \xi_r(\bs_1)\xi_r(\bs_2)\} \psi_r(t_1)\psi_r(t_2),
\end{equation}
where $C(\bs_1,t_1; \bs_2,t_2)$ is the covariance of random field $X(\bs,t) \in L^2(\CS \times \CT)$. 
The covariance (\ref{spatial KL covariance}) greatly simplifies the modeling of space-time interactions.
However, we emphasize that (\ref{spatial KL covariance}) requires that not only $\{\xi_{r}(\bs_i)_{i=1,\dots,N;r\geq 1}\}$ in (\ref{spatial KL}) are uncorrelated at any fixed location $\bs_i$, but more importantly, the uncorrelatedness of spatial random fields \{$\xi_r(\cdot)$\} across $r$, which can not be guaranteed by the KL expansion (\ref{spatial KL}).
Our goal in this article is to introduce a proper concept of separability 
that can be statistically tested in order to determine whether the KL expansion (\ref{spatial KL}) with the simplified covariance (\ref{spatial KL covariance}) is a suitable spatio-temporal model.

We begin with the separability concept in conventional spatio-temporal data analysis, 
which usually assumes the covariance of the spatio-temporal process $X(\bs,t)$, denoted as $C(\bs_1,t_1; \bs_2,t_2)$, can be decomposed as a product of a spatial covariance and a temporal covariance, denoted as $C(\bs_1,\bs_2)$ and $C(t_1,t_2)$, respectively, that is, 
\begin{equation}\label{separability}
 C(\bs_1,t_1; \bs_2,t_2) = C(\bs_1,\bs_2) \, C(t_1,t_2).
\end{equation}
 Separability has been extensively studied in literatures \citep[e.g.][]{sherman2011spatial, cressie2015statistics}.
As the classical spatio-temporal processes are usually non-replicated, i.e., only one realization is observed, the space-time covariance is often estimated using parametric models with a stationary assumption, see \cite{gneiting2006geostatistical} for a class of separable and non-separable covariance functions.
Tests of separability are studied in a number of articles. \cite{mitchell2006likelihood} proposed a likelihood ratio test procedure when replicates from the underlying spatio-temporal process are available. \cite{fuentes2006testing} considered a framework of spectral method in the frequency domain to assess the separability assumption. \cite{li2007A} developed a test based on the asymptotic normality of empirical covariance estimators, which requires stationarity in both space and time domains. 
Recently some nonparametric tests have been developed under the functional data setup \citep{aston2017tests, constantinou2017testing, bagchi2020test}, dealing with two-way functional data with replicates.
In many applications, the null hypothesis of separability is often rejected, see e.g. \cite{fuentes2006testing, mitchell2006likelihood, li2007A, constantinou2017testing}. This indicates that the separability assumption might be too restrictive for space-time correlation structures. Thus, we refer to it as strong separability in the sequel.


To relax the assumption of strong separability, we recall the generalized KL expansion (\ref{spatial KL}) with the cross-covariance (\ref{spatial KL covariance}). 
We highlight that in (\ref{spatial KL}) and (\ref{spatial KL covariance}) we actually assume that the FPC scores \{$\xi_r(\cdot)$\} are uncorrelated spatial random fields for the spatio-temporal process 
$X(\bs,t)$ with mean function $\mu(\bs,t)$ given by
{
$ X(\bs,t) = \mu(\bs,t) + \sum_{r=1}^\infty \xi_r(\bs)\psi_r(t)$.
}
Specifically, through the eigen-decomposition in (\ref{spatial KL covariance}) with a series of eigenfunctions $\{ \psi_r(\cdot) \}$ and the associated eigenvalues being the cross-covariance of spatial random fields \{$\xi_r(\cdot)$\}, i.e., $\E\{ \xi_r(\bs_1)\xi_r(\bs_2)\}$, 
this model views the time domain from a functional perspective while the space domain as random fields. 
In light of this, we also refer to $X(\bs,t)$ as a {\em{spatial functional field}}.
Model (\ref{spatial KL}) highlights the different roles that are played by space and time, distinguishing it from most studies in spatio-temporal statistics or two-way functional data analysis. 
The latter equivalently treats both space and time as functional and does not emphasize the spatial correlation structure. 
Our analysis reveals that strong separability is a special case of this model. We therefore call it {\em{weak separability}}. 


We note that weak separability has been implicitly assumed in many applications 
\citep{li2007nonparametric, ZhouLan2010Rrme, gromenko2012estimation, li2014functional, liu2017functional}. 
However, its validity was rarely examined.
For example, \cite{liu2017functional} proposed a test procedure for strong separability based on a spatial FPCA approach, which implicitly assumed weak separability without justification.
We stress that the expansion (\ref{spatial KL}) only holds pointwisely at each location, but not simultaneously in space and time, i.e., the expansion with uncorrelated spatial processes \{$\xi_r(\cdot)$\} does not hold in general for the spatial functional fields of interest. 
To the best of our knowledge, this is the first work that investigates appropriateness of weak separability and provides a formal test. 
Note that the weak separability defined in \cite{lynch2018test} is for two-way functional data with replicates, which has a different structure from that of our data, and is in fact stronger than our definition (see the derivation in Section \ref{concept of weak separability}). 
Recently \cite{zapata2019partial} introduced a concept of partial separability, which actually extends our definition of weak separability in multivariate functional data. A test of the weak/partial separability assumption, however, is also necessary for the rationality of their work.

We emphasize that our test is proposed for the spatial functional field where only one realization can be observed for $X(\bs,t)$. 
This is common for typical spatio-temporal data, and is distinct from the setup of the replicated (two-way) functional data $\{X_i(\bs,t)\}_{i=1,\dots,N}$ \citep[e.g.][]{constantinou2017testing, lynch2018test, bagchi2020test}, under which replicated realizations  can be observed. 
For such non-replicated data, it is infeasible to estimate the cross-covariance $C(\bs_1,t_1;\bs_2,t_2)$ that plays a critical role for the test of weak separability.
To alleviate this difficulty, we introduce the spatial stationarity for $X(\bs,t)$, as a common practice in spatial statistics \citep[e.g.][]{liu2017functional, hormann2011consistency}, and further utilize the concept of lag covariance function. 
%


The main contribution of this paper is as follows. 
We introduce the definition of weak separability suitable for spatial functional fields, and propose a test procedure with theoretical guarantee for commonly encountered non-replicated spatial temporal process with the aid of spatial stationarity, 
which fills in the need in analyzing spatial functional data. 
In particular, we target at non-replicated spatial functional fields that are intrinsically infinite-dimensional, while the test statistic is based on the empirical FPCA coupled with spatial stationarity that makes estimation feasible. 
The proposed test also distinguishes from approaches with a fixed number of components \citep[e.g.][]{aston2017tests}, allowing truncation to potentially diverge with data size in a nonparametric fashion. 
Based on the asymptotic distribution and appropriately estimated asymptotic covariance, the test is easy to implement, not relying on computationally intensive methods such as bootstrap. 

The rest of the paper is organized as follows. We first introduce the notion of weak separability and suggest appropriate estimation of covariance for spatially stationary functional field in Section \ref{concepts}. The proposed test statistics and their asymptotic properties are presented in Section \ref{test}, which results in a $\chi^2$ test with variance components estimated by parametric/nonparametric methods. We illustrate the empirical performance and validity of the proposed testing procedures by a simulation study in Section \ref{simulation}, and two real data examples in Section \ref{real data}. The article concludes with a discussion in Section \ref{discussion}. Proofs of main theorems are offered in the Appendix, while additional discussions and simulation results are deferred to an online Supplementary Material.

	
	\section{Weak Separability and Covariance Estimation}\label{concepts}
Consider a spatial functional field $X(\bs,t) \in L^2(\CS \times \CT)$, where $\CS \subset \mathbb{R}^2$ is a spatial domain and $\CT$ is a time domain. 
The mean function is $\mu(\bs, t) = \E\{X(\bs,t)\}$ and the covariance function is $C({\bs};t_1,t_2) =\E[\{X(\bs,t_1)-\mu({\bs}, t_1 \} \{X(\bs,t_2)-\mu( {\bs}, t_2 ) \}]$. 
We also define the cross-covariance function 
\begin{equation}\label{cross covariance}
 C(\bs_1,t_1;\bs_2,t_2) = \E \left[ \left\{X(\bs_1,t_1)-\mu (\bs_1, t_1)\} \{X(\bs_2,t_2)-\mu(\bs_2, t_2 )\right\} \right].
\end{equation} 
For a fixed location $\bs$, $X(\bs,t)$ 
is a square integrable function on $\CT$, that is, $X (\bs,\cdot) $ is a random process taking values in $L^2(\CT)$ with the corresponding norm $\|\cdot\|$, which is defined as
$\|X(\bs,\cdot)\| = \left\{ \int_\CT X^2(\bs,t) dt \right\}^{1/2} $.  
Assume that $\E\|X(\bs,\cdot)\|^2<\infty$ for any location $\bs$, 
then the temporal mean and covariance functions, $\mu(\bs; \cdot)$ and $C(\bs;\cdot,\cdot)$, defined on each $\bs$, and the cross-covariance functions $C(\bs_1,\cdot;\bs_2,\cdot)$ between any two locations $\bs_1$ and $\bs_2$, are all well defined with bounded norms \citep{Hsing2015Theoretical, hormann2011consistency}.
We also assume the covariance and cross-covariance are continuous, and $\CS$ and $\CT$ are compact.
These assumptions are used throughout this paper.
The covariance and cross-covariance can also be viewed from the perspective of operators in Hilbert-Schmidt spaces; see \ref{A1} for more details.

 \subsection{Weak separability}\label{concept of weak separability}
In general, a spatial functional field $X(\bs,t)$ can be projected onto some orthonormal basis \{$\psi_r^*(\cdot): r \ge 1$\}, and the projection scores $\xi_{r}^*(\bs)=\int  \{X(\bs,t)-\mu(\bs,t)\} \psi_r^*(t) dt$ are viewed as a series of spatial random fields {in $L^2(\CS)$. This translates into the expansion 
 \begin{equation}\label{weak separability form}
  X(\bs,t) = \mu(\bs,t) + \sum_{r=1}^\infty \xi_r^*(\bs)\psi_r^*(t) 
 \end{equation}
almost surely.
We define that $X(\bs,t)$ is {\em weakly separable} if there exist some basis $\{\psi_r^*(\cdot): r\ge 1\}$, such that for any $r \neq r'$, the scores $\xi_r^*(\cdot)$ and  $\xi_{r'}^*(\cdot)$ are uncorrelated spatial random fields, i.e., $\mbox{cov}\{\xi_r^*(\bs_1),\xi_{r'}^*(\bs_2)\}=0$ for any $\bs_1$ and $\bs_2$ {in $\CS$}.  

{Owing to the uncorrelatedness of FPC scores for any fixed location ${\bs}$, one can show  that the basis $\{\psi_r^*(\cdot)\}$ for a weakly separable $X(\bs,t)$ with the expansion (\ref{weak separability form}) is unique up to a sign change, and 
 consists of the eigenfunctions \{$\psi_r(\cdot)$\} in (\ref{spatial KL covariance}), see Lemma 1 of \citet{lynch2018test}.
Denote $c_r(\bs_1,\bs_2)= \mbox{cov} \{\xi_r(\bs_1), \xi_r(\bs_2)\}$ as the covariance function between $\xi_r(\bs_1)$ and $\xi_r(\bs_2)$, we rewrite (\ref{spatial KL covariance}) as 
\begin{align}\label{cross covariance expansion} 
C(\bs_1,t_1;\bs_2,t_2) & = \E \left[ \left\{ \sum_{r=1}^\infty \xi_r(\bs_1)\psi_r(t_1) \right\} \left\{ \sum_{r=1}^\infty \xi_r(\bs_2)\psi_r(t_2) \right\} \right] \nonumber \\
                                     & = \sum_{r=1}^\infty c_r(\bs_1,\bs_2)\psi_r(t_1)\psi_r(t_2),
\end{align}
which does not have any cross-terms across different $r$ due to the weak separability assumption.} This is vital for the covariance estimation in Section \ref{concept of covariance estimators} and the test of weak separability in Section \ref{test}.
One can see that any process $X(\bs,t)$ satisfying (\ref{cross covariance expansion}) is actually weakly separable according to the definition. 

By contrast, the strong separability with covariance ($\ref{separability}$) is related to the proposed weak separability as follows.
  \begin{pp}\label{weak vs strong} 
        If $X(\bs,t)$ is strongly separable, then it is weakly separable. 
  \end{pp}
  
 
From the cross covariance (\ref{cross covariance expansion}), we can see that a weakly separable $X(\bs,t)$ is also strongly separable if $X(\bs,t)=\xi_1(\bs) \psi_1(t)$. The following proposition provides a sufficient and necessary condition when weak separability can be translated to strong separability. 

 \begin{pp}\label{weak vs strong2} 
Let $X(\bs,t)$ be a weakly separable process with (\ref{weak separability form}) and (\ref{cross covariance expansion}), then it is also strongly separable if and only if there exist a nonnegative decreasing series of $\{\omega_r\}$ with $\sum_{r=1}^\infty \omega_r < \infty$ and a function $\rho( \bs_1,\bs_2 )$ 
such that $c_r(\bs_1,\bs_2) = \omega_r\rho( \bs_1,\bs_2 ) $. 
  \end{pp}


The function $\rho( \bs_1,\bs_2 )$ in Proposition 2 can be understood as a common correlation function for all \{$\xi_r(\cdot)$\}. 
Under weak separability, the spatial correlation functions for different \{$\xi_r(\cdot)$\} can be different, in contrast to a single spatial correlation function under strong separability. 
Meanwhile, weak separability implies a reduced-rank model with truncation of FPCs, thus providing a compromise between strong separability and the general four-dimensional cross-covariance function (\ref{cross covariance}).
One may further consider projection of the random field $\xi_r(\cdot)$ onto a spatial deterministic basis $\{ \zeta_{\varrho} (\cdot)\}_{{\varrho} \ge 1}$ with $ \xi_r(\bs) = \sum_{\varrho=1}^\infty \xi_{r{\varrho}} \zeta_{\varrho} (\bs) $ for each $r$. 
If the random variables \{$\xi_{r{\varrho}}$\} are mutually uncorrelated, then we would have the expansion $X(\bs,t) = \mu(\bs,t) + \sum_{r=1}^\infty \sum_{\varrho=1}^\infty \xi_{r{\varrho}} \zeta_{\varrho}(\bs)\psi_r(t)$ which corresponds to the weak separability defined by \cite{lynch2018test} that is a special case of our definition. 
Hence the proposed weak separability combines the advantages of spatial random fields and functional data analysis, thus leading to a meaningful framework for spatial functional fields.

Note that the form (\ref{cross covariance expansion}) coincides with the coregionalization model in multivariate spatial statistics, which is a generalization of the intrinsic correlation model \citep{li2008testing, sherman2011spatial, li2014functional}. For spatial functional data, strong and weak separabilities can be regarded as the counterparts to the intrinsic and coregionalization models, respectively.

 \subsection{Covariance estimators and eigen-decomposition}\label{concept of covariance estimators}
Covariance function estimation plays an essential role in studying strong/weak separability.
For example, the test statistics in \cite{aston2017tests} and \cite{lynch2018test} are constructed from the full cross and marginal covariance estimators for two-way functional data $X_i(\bs,t)$ with replicates over subjects. For non-replicated spatio-temporal data, only one realization of $X(\bs,t)$ can be observed and thus covariance estimation would be challenging.
In conventional spatio-temporal analysis, assumptions such as stationarity, separability and full symmetry, are usually employed to alleviate the difficulty in covariance estimation \citep{gneiting2006geostatistical}.
The proposed weak separability, as we mentioned before, treats space and time from different perspectives, which motivates us to estimate the covariance function across space but not across time.
Similar to \cite{hormann2011consistency}, a reasonable assumption is that all locations share a common mean function and a common covariance function, i.e.
\begin{equation}\label{assump_stationarity1}
\mu({\bs},t) = \mu(t), \quad
C(\bs,t_1;\bs,t_2)=C(t_1,t_2).
\end{equation}

Suppose $X(\bs_i,t)$ is observed across $N$ spatial locations, $i=1,\dots,N$. For ease of notation, we also denote $X(\bs_i,t)$ as $X_i(t)$.
According to (\ref{assump_stationarity1}), we could estimate $\mu(t)$ by the sample mean $\hat\mu (t)=1/N\sum_{i=1}^N X_i(t)$, and estimate $C(t_1,t_2)$ by the sample covariance function 
\begin{equation*}\label{empirical covariance function} 
 \wh{C}(t_1,t_2)=\frac{1}{N}\sum_{i=1}^N \left\{ X_i(t_1)-\hat \mu(t_1) \right\} \left\{ X_ i(t_2)-\hat \mu(t_2) \right\}.
\end{equation*}
With this estimated covariance function, one could perform the standard FPCA, which, however, is not useful for studying the weak separability of interest. This is due to the degenerate moment estimates of the correlation of FPC scores, which we shall elucidate in Section \ref{test statistics}. 

As the covariance function in (\ref{assump_stationarity1}) does not contain information across spatial locations, it is necessary to consider the cross-covariance function defined in (\ref{cross covariance}). To make the estimation feasible for non-replicated spatial functional field, we assume that $X(\bs,t)$ is second-order stationary spatially, that is, for some function $C^{(\bh)}(t_1,t_2)$, where $\bh=\bs_1-\bs_2$ is a spatial lag, 
\begin{equation}\label{assump_stationarity2}
C(\bs_1,t_1;\bs_2,t_2)=C^{(\bh)}(t_1,t_2). 
\end{equation} 
We will refer to $C^{(\bh)}(t_1,t_2)$ as the {\em lag covariance}, which can be estimated by  
\begin{equation}\label{empirical lag covariance function}
\wh C^{(\bh)}(t_1,t_2)=\frac{1}{N_{\bh} }\sum_{\bs_i-\bs_{i'}=\bh}\left\{X_i(t_1)-\hat \mu(t_1)\right\}\left\{X_ {i'}(t_2)-\hat \mu(t_2)\right\},
\end{equation}
where $N_{\bh}$ is the number of pairs $(i,i')$ satisfying $\bs_i-\bs_{i'}=\bh$.
This lag covariance estimator enables us to borrow information spatially, and plays a crucial role in the proposed test. 

Note that the assumed spatial stationarity implies the stationarity of the spatial random fields \{$\xi_{r}(\cdot)\}$. 
We stress that weak separability (or strong separability) is a separate issue from spatial stationarity; the latter is assumed primarily due to lack of replicates for estimating the cross-covariance. 
One may consider other structures that facilitate borrowing information spatially, such as local stationarity \citep{hormann2010weakly}. To better focus on weak separability we do not consider such complex structures here. 
For a comprehensive understanding, we provide more discussion about the extension of the weak separability test to (non-stationary) replicated data, and the sensitivity analysis about stationarity in Section S.1 of the Supplementary Material.

For a weakly separable and spatially stationary $X(\bs,t)$ with the expansion (\ref{weak separability form}), let $\omega_r=\E \{ \xi_r^2(\bs) \} $ and $\eta_r^{(\bh)}=\E \{ \xi_r(\bs) \xi_r(\bs + \bh) \} $ for $r=1,2,\dots$, 
the cross covariance (\ref{cross covariance expansion}) leads to
$ C(t_1,t_2) = \sum_{r=1}^\infty\omega_r \psi_r(t_1)\psi_r(t_2) $, and more importantly, 
\begin{equation}\label{lag covariance eigen-decomposition}
 C^{(\bh)}(t_1,t_2) = \sum_{r=1}^\infty\eta_r^{(\bh)} \psi_r(t_1)\psi_r(t_2),
 \end{equation}
It is interesting to note that the lag covariance function can be decomposed 
with the same set of eigenfunctions 
but different eigenvalues. 
Since the proposed test is mainly based on the lag covariance, we assume  $\eta_1^{(\bh)}>\eta_2^{(\bh)}>\cdots \geq 0$,  noting that the ordering of eigenfunctions \{$\psi_r(\cdot):r=1,2,\dots$\} now may be slightly different from that in a standard FPCA requiring decreasing \{$\omega_r: r=1,2,\dots$\}, and also slightly different for different $\bh$.   
{This assumption is reasonable and can be satisfied by smooth stationary spatial random fields. For instance, if \{$\xi_r(\cdot)$\} are mean-squared continuous, that is, $ \lim_{\bh\rightarrow 0} \E \{\xi_r(\bs + \bh ) - \xi_r(\bs) \}^2 = 0$, then $\eta_r^{(\bh)}$ are close to $w_r$ in some  neighborhood of $\bh$. Moreover, correlation among nearby locations (i.e., small $|\bh|$) is typically of interest for examining weak separability in spatial statistics.
In practice it is easy to arrange $\{\omega_r\}$ corresponding to \{$\psi_r(\cdot):r=1,2,\dots$\} by matching their estimates, while small lags are recommended for the proposed test; see Section \ref{implementation} for more discussion.}

Next we estimate the eigen-elements \{$\psi_r(\cdot)$\}, \{$\omega_r$\} and \{$\eta_r^{(\bh)}$\} 
by solving the eigen-equations 
\begin{equation}\label{eigen-decomposition}
\int \wh{C}(t_1,t_2)\hat\psi_{r}(t_1)dt_1=\hat\omega_{r}\hat\psi_{r}(t_2), {\quad \mbox{ for } r = 1,2\dots,}
\end{equation}
and
\begin{equation}\label{eigen-decomposition lag}
\int \wh C^{(\bh)}(t_1,t_2)\hat\psi_{r}^{(\bh)}(t_1)dt_1=\hat\eta_{r}^{(\bh)} \hat\psi_{r}^{(\bh)}(t_2), {\quad \mbox{ for } r = 1,2\dots.}
\end{equation}
Note that both $\hat\psi_r(\cdot)$ and $\hat\psi_r^{(\bh)}(\cdot)$ are consistent estimates of $\psi_r(\cdot)$, while $\hat\psi_r^{(\bh)}(\cdot)$ is associated with $\hat\eta_{r}^{(\bh)}$  which contains information on the cross-covariance. In other words, the decomposition (\ref{lag covariance eigen-decomposition}) is a result of weak separability using the lag covariance.
In contrast to $C(t_1,t_2)$, the lag covariance $C^{(\bh)}(t_1,t_2)$ contains covariance information across spatial locations, which is critical for the purpose of testing weak separability.

The proposed lag covariance estimation is developed for spatially stationary $X(\bs,t)$. 
For ease of expression, in what follows we present our methods based on the isotropic lag covariance $C^{(h)}(t_1,t_2)$, 
where $h=|\bs_1-\bs_{2}|$ is the Euclidean distance between $\bs_1$ and $\bs_{2}$, instead of the directional lag $\bh$ in (\ref{assump_stationarity2}). We estimate $C^{(h)}(t_1,t_2)$ by $\wh C^{(h)}(t_1,t_2)$, which is obtained by substituting $\bs_i-\bs_{i'}=\bh$ with $| \bs_i-\bs_{i'} |=h$ and $N_\bh$ with $N_h$ (i.e. the number pairs separated by a distance $h$) in (\ref{empirical lag covariance function}). A discussion on how to choose $h$ is given in Section \ref{implementation}. For regularly spaced data, one can assume {$N_h \asymp N$} 
under a suitable choice of $h$, {where $a_n \asymp b_n$ means $0< \lim \inf a_n/b_n < \lim \sup a_n/b_n<\infty$ for any positive sequences \{$a_n$\} and \{$b_n$\}. }
This ensures that $\wh C^{(h)}(t_1,t_2)$ has the same convergence rate as $\wh C(t_1,t_2)$.

\begin{remark}\label{nongrid}
 \rm For irregularly spaced data, there may be few pairs that are separated exactly by a given lag distance $h$. In this case, we instead use a kernel smoothing type of estimator by including more pairs with distances within a small neighborhood around $h$. Specifically, let $h_{i,i'}= |\bs_i - \bs_{i'}|$, and let $\kappa_\delta(\cdot)$ be a univariate kernel function with bandwidth $\delta$, we consider
$\wh C_{\delta}(h;t_1,t_2)= \sum_{i, i'} { \left\{X_i(t_1)-\hat\mu(t_1)\right\}\left\{X_ {i'}(t_2)-\hat\mu(t_2)\right\} \kappa_\delta( h - h_{i,i'} )  }  / { \sum_{i, i'} \kappa_\delta( h - h_{i,i'} ) }$. Since our data examples involve only regularly spaced data, we do not pursue further in this direction.
 \end{remark}

\section{Test of Weak Separability }\label{test}
            
\subsection{ Statistic based on lag covariance}\label{test statistics}      
According to Section \ref{concepts}, a spatial functional field $X(\bs,t)$ satisfying (\ref{assump_stationarity1})  can be projected onto a unique orthogonal basis system \{$\psi_r(\cdot)$\} from the covariance $C(t_1,t_2)$, resulting in a series of spatial random fields \{$\xi_r(\cdot)$\}. 
We stress again that, owing to the spatial correlation among different locations, a rigorous test for weak separability is necessary, i.e., whether $\cov\{\xi_j(\cdot),\xi_k(\cdot) \}=0$ for any $j\not=k$. 

Testing the correlation of two spatial processes is not a new question. \cite{clifford1989assessing} and \cite{dutilleul1993modifying} presented a modified $t$-test. \cite{gromenko2012estimation} proposed a test for the correlation of two spatial fields using the estimated FPC scores, while its asymptotic distribution is based on the true scores.
We shall show in Section \ref{properties} that, although the spatial FPC scores can be consistently estimated, the test statistic using the estimated FPC scores has a different distribution from that of its counterpart using true FPC scores.
Suppose that $X(\bs,t)$ is weakly separable and we have the observation $X(\bs_i,t)$, using (\ref{weak separability form}), 
\begin{equation*}\label{empirical weak separability form}
 X(\bs_i,t) = \mu(t) + \sum_{r=1}^\infty \xi_r(\bs_i)\psi_r(t),
 \end{equation*}
where the true scores \{$\xi_r(\bs_i)$\} are not observed and need to be estimated. For ease of presentation, we write $\xi_r (\bs_i)$ as $\xi_{ir}$.
A straightforward method is to perform eigen-decomposition of the empirical covariance $\wh{C}(t_1,t_2)$ by (\ref{eigen-decomposition}) 
and estimate $\xi_{ir}$ by $\hat\xi_{ir}=\int \{X_i(t)-\hat \mu (t) \}\hat\psi_{r}(t)dt$. Then a naive test statistic can be defined as $T_0(j,k)=N^{-1/2}\sum_{i=1}^N\hat{\xi}_{ij}\hat{\xi}_{ik}$, for $j\not=k$. However, it can be easily shown that $T_0(j,k)$ is degenerate, by noting
\begin{align}\label{degeneration derivation}
T_0(j,k) &=N^{-1/2}\sum_{i=1}^N\int\{X_i(t)-\hat\mu(t)\}\hat\psi_j(t)dt\int\{X_i(t)-\hat\mu(t)\}\hat\psi_k(t)dt \nonumber \\
                                                                             &= \sqrt{N} \iint {N}^{-1} \sum_{i=1}^N\{X_i(t_1)-\hat\mu(t_1)\}\{X_i(t_2)-\hat\mu(t_2)\}\hat\psi_j(t_1)\hat\psi_k(t_2)dt_1dt_2 \nonumber\\
                                                                             &= \sqrt{N} \iint\widehat C(t_1,t_2) \hat\psi_j(t_1)\hat\psi_k(t_2) dt_1dt_2 = 0,
\end{align}
{since $\hat\psi_j(t)$ and $\hat\psi_k(t)$ are orthogonal eigenfunctions of $\widehat C(t_1,t_2)$}. 
A different approach is to consider 
\begin{equation}\label{projected score}
\hat\xi_{ir}^{(h)}=\int \{X_i(t)-\hat \mu (t) \}\hat\psi_{r}^{(h)}(t)dt,
\end{equation}
 where \{$\hat\psi_{r}^{(h)} (\cdot)$\} are the eigenfunctions obtained by solving (\ref{eigen-decomposition lag}), with the empirical lag covariance $\wh C^{(\bh)}(t_1,t_2)$ therein being replaced by its isotropic counterpart $\wh C^{(h)}(t_1,t_2)$. Then for $j\not=k$, we define the statistic as follows,
\begin{equation}\label{definition of Tn}
T_h(j,k)=\frac{1}{\sqrt{N}}\sum_{i=1}^N \hat\xi^{(h)}_{ij}\hat\xi^{(h)}_{ik}.
\end{equation}
By a similar derivation as (\ref{degeneration derivation}), we can show $T_h(j,k)=  \sqrt{N} \iint\widehat C(t_1,t_2) \hat\psi_j^{(h)}(t_1)\hat\psi_k^{(h)}(t_2) dt_1dt_2 $, and
the degeneration would then no longer occur.
We will study the asymptotic behavior of $T_h(j,k)$ based on which we develop our test for weak separability.

        \subsection{Asymptotic properties}\label{properties}      

We study asymptotic properties of (\ref{definition of Tn}) under an {increasing domain} setting similar to that used in \cite{li2014functional, zhang2020unified}. Consider a sequence of spatial domains $\CS_N$ with expanding areas, while the time domain $\CT$ remains fixed. Specifically, we assume that
\begin{equation}\label{increasing domain}
C_1 N \leq |\CS_N| \leq C_2 N,\mbox{ and } C_1 \sqrt{N} \leq|\partial \CS_N| \leq C_2 \sqrt{N} \mbox{ for some }C_1, C_2>0,
\end{equation}
where $|\CS_N|$ denotes the area of $\CS_N$, $\partial \CS_n$ the boundary of $\CS_N$ and $|\partial \CS_N|$ the perimeter of $\partial \CS_N$. 
This condition basically says that the spatial domain increases in all directions, and its shape is not too irregular. It also satisfies the definition of Type C sampling scheme in \cite{hormann2011consistency}, which is required for the consistency of the mean and covariance estimators.

{To investigate the limit distribution of $T_h(j,k)$, we impose the following regularity conditions for the moment of $X(\bs,t)$.
}
{
\begin{asmp}\label{condition: high order moment}
\rm For any location $\bs \in \CS$ and each $C>0$, there exists an $\epsilon>0$ such that
\begin{equation}\label{smoothness}
\sup_{t\in \CT} \E \{ | X(\bs, t) |^C \} < \infty, \mbox{ and } \sup_{t_1,t_2\in \CT} \E [ \{ |t_1-t_2|^{-\epsilon} | X(\bs,t_1) -X(\bs,t_2) | \}^C ] < \infty;
\end{equation} 
\begin{equation}\label{higher moment bound}
\mbox{for each integer } b \geq 1, \omega_r^{-b} \, \E \{ \xi_r(\bs)^{2b}  \} \mbox{ is bounded uniformly in }r ;
\end{equation} 
\begin{equation}\label{fourth moment}
\E \| X(\bs,\cdot) \|^v<\infty \mbox{ for some } v>4;
\end{equation} 
\end{asmp}
Conditions (\ref{smoothness}) and (\ref{higher moment bound}) consist of regular assumptions for functional data \citep[see e.g.][]{hall2006properties, kong2016partially}; for example, a Gaussian process with H\"{o}lder continuous sample paths satisfies (\ref{smoothness}) and (\ref{higher moment bound}). 
Condition (\ref{fourth moment}) imposes the boundedness of the high-order moment, which is required for the central limit theorem (CLT) of the covariance estimators.
A standard conclusion in Hilbert space is that the CLT for the covariance constructed by i.i.d random elements holds if the fourth-order moment of the process is bounded,
while Condition (\ref{fourth moment}) requires a slightly stronger condition on the moment of $X(\bs,\cdot)$ due to the spatial correlation.
}

To provide the asymptotic normality for the covariance and lag covariance of the spatial functional field, 
we then introduce the following strong mixing coefficient \citep{rosenblatt1956central}:
\begin{equation}\label{strong mixing coefficient}
\alpha_{X}(u) \equiv \sup_{\substack{ E_1,E_2\subset \mathbb{R}^2 \\ d(E_1,E_2)\geq u} } \sup_{\substack{ A_1\in \mathscr{F}_X(E_1), \\ A_2\in \mathscr{F}_X(E_2) } }   | \BP(A_1\cap A_2)-\BP(A_1)\BP(A_2)| ,
\end{equation}
where 
$\mathscr{F}_X(E)$ represents the $\sigma$-algebra generated by \{$X(\bs,t): (\bs,t) \in E \times \CT$\} for any $E \subset \mathbb{R}^2$ \citep{zhang2020unified}, $d(E_1,E_2)= \inf\{d(\rho_1,\rho_2):\rho_1\in E_1,\rho_2\in E_2,  E_1,E_2\subset \mathbb{R}^2 \} $ denotes the minimal Euclidean distance between $E_1$ and $E_2$.
The mixing coefficient quantifies the spatial dependence of the random processes at different locations.
One can see that if the observations are spatially independent, then $\alpha_{X}(u)=0$ for all $u>0$.
For a $m$-dependent random field (i.e. observations are independent if their distance is larger than $m$), $\alpha_{X}(u)$ is equals to $0$ for $u\geq m$. 
Following \cite{zhang2020unified}, we then require 
\begin{asmp}\label{condition: mixing}
\rm $X(\bs,t)$ is strictly stationary in $\bs$ with the mixing coefficient defined in (\ref{strong mixing coefficient}), and for $v$ in (\ref{fourth moment}), there exists $\beta>v/(v-4)$ such that
\begin{equation}\label{decay on alpha}
\alpha_X(u)=O(u^{-\beta}). 
\end{equation}
\end{asmp}

Condition \ref{condition: mixing} says that as $u$ increases, the spatial dependency in $X(\bs,t)$ decreases at a polynomial rate in $u$. 
As pointed out by \cite{sherman2011spatial}, Condition \ref{condition: high order moment} and \ref{condition: mixing} actually provide a tradeoff between the strength of spatial correlation (quantified by the mixing condition) and the heaviness of the tails of the distribution (quantified by the number of moments), which guarantees the CLT for $\wh C(t_1,t_2)$ and $\wh C^{(h)}(t_1,t_2)$ in Lemmas \ref{asymptotic normality of iid covariance operators} and \ref{asymptotic normality of covariance operators} of \ref{A3}.
Note that Condition \ref{condition: mixing} with $\alpha$-mixing coefficient (\ref{strong mixing coefficient}) provides a sufficient condition for the asymptotic normality, which may also hold under different mixings \citep[see e.g.][]{bosq2012nonparametric}.



As discussed in Section \ref{concept of covariance estimators}, for a weakly separable and spatially stationary $X(\bs,t)$, the lag covariance function $C^{(h)}(t_1,t_2)$ can be decomposed with the same set of eigenfunctions \{$\psi_r(\cdot)$\} {as} $C(t_1,t_2)$ but different eigenvalues \{$\eta_r^{(h)}$\}. 
The next condition is to control the decay rates of the eigenvalues as the standard FPCA approach. 
As the test statistic (\ref{definition of Tn}) is constructed based on the lag covariance, we only need to impose the decay condition on the eigenvalues \{$\eta_r^{(h)}$\}. 
{\begin{asmp}\label{condition: eigenvalues}
 \rm There exists $a>1$, $C>0$, s.t. $\eta_r^{(h)}-\eta_{r+1}^{(h)} \ge C r^{-a-1}$ for $r\ge1$ in some neighborhood of $h$ including $0$.
\end{asmp}}

This condition is similar to that adopted by \cite{hall2007methodology, kong2016partially}, and implies that $\eta_r^{(h)}\ge Cr^{-a}$ with $a>1$ due to boundedness.



For infinite-dimensional functional processes,  truncation is usually applied to control complexity of the approximation to the underlying process. 
With the truncated form denoted by $X_{R_N}(\bs,t) = \mu(t) + \sum_{r=1}^{R_N} \xi_r(\bs)\psi_r(t)$,
it is important to control $R_N$ appropriately. 
The following condition says that $R_N$ cannot be too large due to the increasingly unstable FPC estimates. 
\begin{asmp}\label{condition: truncation}
{$R_N^{2a+4}/N=o(1)$. }
\end{asmp}

{
 \begin{thm}\label{asymptotic express of Tn}
Consider a spatially stationary functional field $X(\bs,t)$ satisfying (\ref{assump_stationarity1}) and (\ref{assump_stationarity2}), observed in an increasing domain scheme (\ref{increasing domain}) with a finite $N_h$.
Let 
 \begin{equation}\label{true term}
 T_h^*(j,k) = \frac{1}{\sqrt{N}}\sum_{i=1}^N \xi_{ij}\xi_{ik} - \frac{\sqrt{N}}{\rho_{jk}^{(h)} N_h}\sum_{|\bs_i-\bs_i'| = h } \xi_{ij}\xi_{i'k},
 \end{equation}
  where 
 \begin{equation}\label{rho_jk}
\rho_{jk}^{(h)}=(\eta_j^{(h)}-\eta_k^{(h)})(\omega_j-\omega_k)^{-1}.
 \end{equation}
Assume that $c\leq \rho_{jk}^{(h)}\leq C$ for positive constants $c$ and $C$, and Conditions \ref{condition: high order moment}-\ref{condition: truncation} hold. If $X(\bs,t)$ is weakly separable, then on an event set $\CE_{N,R_N}$ defined in (\ref{high probability set}) with $P(\CE_{N,R_N})\rightarrow 1$, we have 
 \begin{equation*}
\E \, | T_h(j,k) -  T_h^*(j,k) |^2 = O(R_N^{2a+4}N^{-1})
\end{equation*}
uniformly in $j,k=1,\dots,R_N, j\not=k$.
\end{thm}
}
Note that the first term in (\ref{true term}) is the counterpart to $T_h$ using the true scores, while the second term is the non-negligible difference between them given $N_h \asymp N $. The proof of Theorem \ref{asymptotic express of Tn} is deferred to \ref{A3}.

        \subsection{ Proposed $\chi^2$ test }\label{test procedure} 
        
Theorem \ref{asymptotic express of Tn} presents the asymptotic expression of $T_h(j, k)$ for any pair $(j,k)$, $j,k\le R_N$, where $R_N$ may diverge with $N$.
Denote the index set $\{ (j,k): j,k=1,\dots,R_N, j < k\}$ by $\mathbb{I}_{R_N}$ with the cardinality $R_N^*=R_N(R_N-1)/2$.
{
Let $T_h(\mathbb{I}_{R_N})$ be the vector by stacking $\{T_h(j, k): (j, k)\in \mathbb{I}_{R_N}\}$, and correspondingly $T^*_h(\mathbb{I}_{R_N}) = \{T_h^*(j, k): (j, k)\in \mathbb{I}_{R_N}\} $.
We shall present the convergence rate on the full set of $\mathbb{I}_{R_N}$, then the result for the subsets follows immediately.
}

{Theorem \ref{joint asymptotic normality of Tn} will consider the probability measure between $T_h(\mathbb{I}_{R_N})$ and $T^*_h(\mathbb{I}_{R_N})$. 
We shall show that $\|T_h(\mathbb{I}_{R_N})-T^*_h(\mathbb{I}_{R_N})\| = o_p(1)$, with a slightly stronger assumption than Condition \ref{condition: truncation}
which is graphically illustrated in Figure \ref{truancation_map}.
}

\begin{cond}{condition: truncation}
{$ R_N^{(2a+6)} / N=o(1)$.}
\end{cond}

        \begin{figure}[h!]
                         \centering
        			\caption{An illustration of Conditions \ref{condition: truncation} and \ref{condition: truncation}$^*$. The graph below shows the relationship between the eigenvalue decay rate $a$ ($x$-axis) and the polynomial order of truncation $R_N$ ($y$-axis). The outer shaded area, which is surrounded by $a>1$ (solid line) and {$y> 2a+4$} (dashed line), corresponds to Condition \ref{condition: truncation}, while the inner shaded area (filled by dashes) surrounded by $a>1$ and {$y> 2a+6$} (dash-dotted line), corresponds to Condition \ref{condition: truncation}$^*$.}
        			\includegraphics[width=0.6\linewidth]{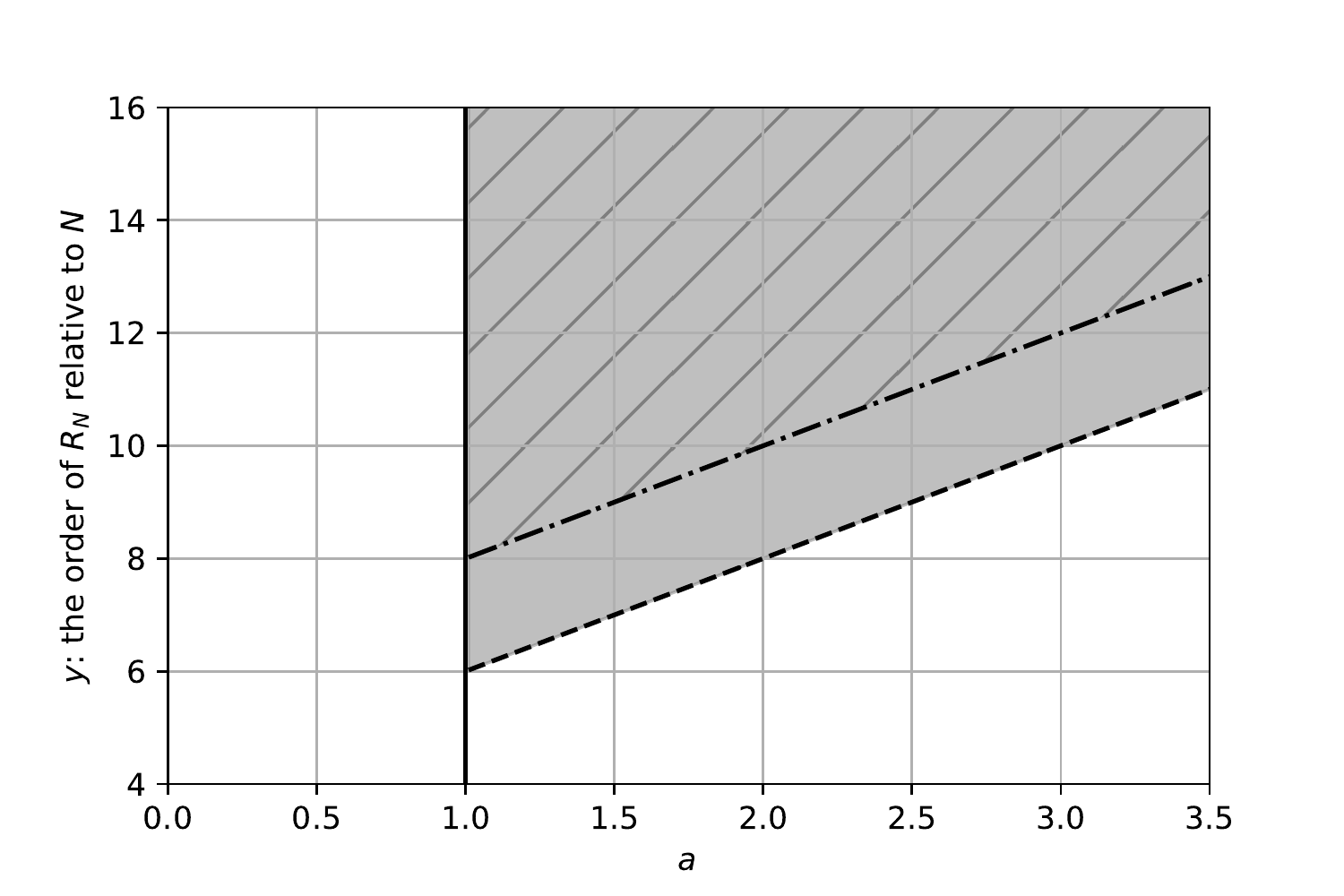}
        			\label{truancation_map}
        	\end{figure}

\begin{remark}\label{diversion}
 \rm Here we consider possibly diverging $R_N$ (thus $R_N^*$) to avoid missing potential signal in high-order terms,
different from the tests in \cite{aston2017tests, constantinou2017testing, lynch2018test} which considered finite truncation. Our proposal allows the truncation to diverge with the sample size, which is integrated into the limiting distribution and reflects the nonparametric nature of the test.
\end{remark}

To describe the joint distribution of $T_h(\mathbb{I}_{R_N})$, of which the dimension $R_N^*$ diverges with $N$, we introduce the following Prokhorov metric $\pi$: 
$$ \pi(\mu,\nu) \equiv \inf \{ \varepsilon \geq 0 : \mu (A)\leq \nu(A^\varepsilon) + \varepsilon \mbox{ and }  \nu (A)\leq \mu(A^\varepsilon) + \varepsilon \mbox{ for all } A\in \mathscr{B}(M) \},$$
where $(M,d_M)$ is a metric space with its Borel sigma algebra $\mathscr{B}(M)$, $\mu$ and $\nu$ are two probability measures on the measure space $(M,\mathscr{B}(M))$, $A^\varepsilon= \{x:\exists y\in A, d_M(x,y) < \varepsilon \} $ is the $\varepsilon$-neighborhood of $A$ \citep[e.g.,][Section 6 of Chapter 1]{billingsley1999convergence}.

\begin{thm}\label{joint asymptotic normality of Tn} 
Assume that the conditions of Theorem \ref{asymptotic express of Tn} and Condition {\ref{condition: truncation}$^*$}  hold, and $N_h \asymp N $. If $X(\bs,t)$ is weakly separable, then
$$ \pi \left( T_h(\mathbb{I}_{R_N} \right) , \CZ_{R_N^*} ) \longrightarrow 0, $$
where $\CZ_{R_N^*} \sim N_{R_N^*}(0,\Gamma)$ is a $R_N^*$-dimensional Gaussian random vector with mean zero and covariance matrix $\Gamma$ given in (\ref{asym covariance}) in \ref{proof2}, 
and $\pi(\cdot, \cdot)$ is the Prokhorov metric. 
\end{thm}

To derive our test statistic for weak separability, we need to estimate $\Gamma$, which, according to the explicit expression (\ref{asym covariance}) in \ref{proof2}, relies on the cross fourth-order moments of the FPC scores, i.e., $\E ( \xi_{i_1 j}\xi_{i_2 k} \xi_{i_3 j'} \xi_{i_4 k'} )$, $(j, k),(j', k')\in \mathbb{I}_{R_N}$ and $i_1,i_2,i_3,i_4=1,\dots,N$. 
Note that Condition \ref{condition: high order moment} implies the boundedness of these cross fourth-order moments for a weakly separable process. 
While the assumed second-order stationarity for the projected spatial random fields \{$\xi_r(\cdot)$\} makes it possible to estimate the second-order moments by borrowing information spatially, additional assumptions would be needed in order to estimate these higher-order moments. In this paper, we assume that \{$\xi_r(\cdot)$\} are Gaussian. We note that Gaussian random fields have been widely studied in spatial statistics and that similar assumptions have been also made in the functional data setting \citep{aston2017tests}. 
We also provide more discussion about the Gaussian assumption with some sensitivity analysis and numerical results in Section S.2 of the Supplementary Material.

 Under the assumed Gaussionality, the projected random fields \{$\xi_r(\cdot)$\} would be jointly independent across $r$. Consequently the cross fourth moments $\E ( \xi_{i_1 j}\xi_{i_2 k} \xi_{i_3 j'} \xi_{i_4 k'} ) $ become 0 for $j\not=j'$ or $k\not=k'$, 
and the off-diagonal elements $\mbox{cov}\{T_h(j,k),T_h(j',k')\}$ based on $\E ( \xi_{i_1 j}\xi_{i_2 k} \xi_{i_3 j'} \xi_{i_4 k'} ) $ would all be 0. 
For the diagonal elements, the cross moments $\E ( \xi_{i_1 j}\xi_{i_2 k} \xi_{i_3 j} \xi_{i_4 k} ) $ can also be expressed by the products of $\E ( \xi_{i_1 j}\xi_{i_3 j} )$ and $\E (\xi_{i_2 k} \xi_{i_4 k} )$ due to independence. 
According to Theorem \ref{joint asymptotic normality of Tn}, the vector $T_h(\mathbb{I}_{R_N})$ can be approximated by a $R_N^*$-dimensional multivariate Gaussian distribution.
We then formulate the proposed $\chi^2$ test statistic as 
\begin{equation}\label{chi2 test statistic}
 S_h( \mathbb{I}_{R_N} ) =\sum_{(j,k) \in  \mathbb{I}_{R_N} } \{T_h(j,k)/\sigma_{j,k}\}^2,
\end{equation} 
where \{$\sigma_{j,k}^2$\} are the diagonal elements of $\Gamma$ as in (\ref{sigma^2(j,k)}).

 \begin{cl}\label{asymptotic under gaussian}  
 	Under the conditions of Theorem \ref{joint asymptotic normality of Tn}, and the 
	additional assumption that $X(\bs,t)$ is Gaussian, the covariance matrix $\Gamma$ in Theorem \ref{joint asymptotic normality of Tn} becomes diagnal,  $\Gamma = \mbox{diag} \{\sigma^2_{j, k}: (j, k)\in \mathbb{I}_{R_N}\} $ with the elements 
\begin{equation}\label{sigma^2(j,k)}
	\sigma^2_{j,k}= \frac{1}{N} \mbox{tr}(U_jU_k) + \frac{N}{(\rho_{jk}^{(h)} N_h)^2} \mbox{tr}(U_{j,1}U_{k,2}) - \frac{2}{\rho_{jk}^{(h)} N_h} \mbox{tr}(V_{j,1}V^{\trans}_{k,2}),
\end{equation}
where $\rho_{jk}^{(h)}$ is defined in (\ref{rho_jk}), 
$U_r = \E(\bm\xi_r\bm\xi_r^{\trans})$ is the covariance matrix of $\bm\xi_r=(\xi_{1r},\dots,\xi_{Nr})^{\trans}$, and $U_{r,1},U_{r,2}, V_{r,1},V_{r,2}$ are covariance matrices based on $U_r$ 
with explicit formulas given in (\ref{covariance notation}) in \ref{proof2}. 
Moreover, we have 
$$\pi \left( S_h(\mathbb{I}_{R_N}), \chi^2_{R_N^*} \right)  \longrightarrow 0,$$
where $ \chi^2_{R_N^*}$ is a chi-squared distribution with $R_N^*=R_N(R_N-1)/2$ degrees of freedom, and $\pi(\cdot, \cdot)$ is the Prokhorov metric.   \end{cl}

\subsection{ Implementation and parameter selection}\label{implementation}

According to Corollary \ref{asymptotic under gaussian}, we can perform a $\chi^2$ test based on $S_h( \mathbb{I}_{R_N} )$, with an estimate of $\Gamma$ and a suitable truncation $R_N$. For the choice of $R_N$, Conditions \ref{condition: truncation} and  {\ref{condition: truncation}$^*$} provide merely theoretical magnitudes. On the other hand, $R_N$ cannot be too small, so that under the alternative hypothesis the signal for $\{\xi_r:r\le R_N\}$ can be detected.
In practice, we suggest to conduct the proposed test over a range of $R_N$, e.g. as determined by the fraction of variance explained (FVE),  $\mbox{FVE} (R_N) = \{ \sum_{r=1}^{R_N} \hat\eta_r^{(h)} \} / ( \sum_{r=1}^{\infty} \hat\eta_r^{(h)} ) $ for FVE=80\%, 90\% or 95\% etc. 
Here we use the eigenvalues $\{\hat\eta_r^{(h)}\}$ calculated from the empirical lag covariance $\wh C^{(h)}(t_1,t_2)$, instead of $\{\hat\omega_r\}$ from the empirical covariance function $\wh C(t_1,t_2)$, as the statistic $T_h(j,k)$ is constructed based on the lag covariance. 
Under the null hypothesis, according to the definition of weak separability, the test results should agree when using different values of $R_N$. 
On the other hand, 
when $X(\bs,t)$ is not weakly separable, $\xi_j(\cdot)$ and $\xi_k(\cdot)$ are correlated for some $(j,k)$ and the corresponding $T_h(j,k)$ may be large. Therefore, an appropriate $R_N$ should include all correlated FPC fields and a reliable conclusion can be made with agreements across different $R_N$ values. 
This is demonstrated in the data applications in Section \ref{real data}, and more discussion and simulation results about the choice of $R_N$ are given in Section S.3.1 of the Supplementary Material.

For the estimation of the covariance matrix $\Gamma$, 
according to (\ref{sigma^2(j,k)}) in Corollary \ref{asymptotic under gaussian}, 
we need to estimate \{$\rho_{jk}^{(h)}: (j, k) \in \mathbb{I}_{R_N}$\} and the covariance matrices \{$U_r: r=1,\dots,R_N$\}. 
To estimate $\rho_{jk}^{(h)}$ defined in (\ref{rho_jk}), we need to estimate two sets of eigenvalues \{$\hat\omega_r$\} and \{$\hat\eta_r^{(h)}$\} which could be obtained 
by eigen-decompositions of the sample covariance $\wh{C}(t_1,t_2)$ and the lag covariance $\wh C^{(h)}(t_1,t_2)$, respectively.
In practice we first obtain \{$\hat\eta_r^{(h)}, \hat\psi_r^{(h)}$\} from $\wh{C}(h;t_1,t_2)$ for $r=1,2,\dots,R_N$  with $\hat\eta_1^{(h)}>\hat\eta_2^{(h)}>\dots \ge 0$. The corresponding \{$\hat\omega_r$\} can be determined by simply matching the estimated eigenfunctions of $\wh{C}$, denoted by \{$\hat\psi_{r'}: r'\geq1$\}, to \{$\hat\psi_r^{(h)}: r\ge 1$\} by $\hat\psi_r = \arg \max_{\psi_{r'}}\{| \int \hat\psi_r^{(h)}(t) \hat\psi_{r'}(t) dt | :r'\geq 1\} $, which performs well in our numerical studies.
We then estimate $\rho_{jk}^{(h)}$ by $\hat\rho_{jk}^{(h)}=({\hat\eta_j^{(h)}-\hat\eta_k^{(h)}})/({\hat\omega_j-\wh\omega_k})$ for each $(j,k)$.

For the estimation of covariance matrices \{$U_r: r=1,\dots,R_N$\}, we first note that the diagonal elements in $U_r$ are the variance of $\xi_{r}(\cdot)$, thus can be estimated by $\hat\omega_r$. 
To estimate the off-diagonal elements in $U_r$, i.e. the cross-covariance $\E \{\xi_{r}(\bs_i)\xi_{r}(\bs_{i'} ) \} $ when $\bs_i \not= \bs_{i'}$, note that under the spatial stationarity we can derive 
\begin{equation}
\E \{\xi_{r}(\bs_i)\xi_{r}(\bs_{i'} ) \} = \omega_r \rho_r(|\bs_i-\bs_{i'}|),
\end{equation}
where $\rho_r(\cdot)$ is a correlation function that could vary with $r$. 
For each pairwise distance $d$ in the set $\CD=\{|\bs_i-\bs_{i'}|: \bs_i \not= \bs_{i'}, \bs_i \in \CS,\bs_{i'} \in \CS\}$, we estimate $\rho_r(d)$ by $\tilde\rho_r(d)= ({N_d}^{-1}\sum_{|\bs_i-\bs_{i'}|=d}\hat\xi_{ir}\hat\xi_{i'r})/\hat{\omega}_r$, where $N_d$ is the total number of pairs ($\bs_i, \bs_{i'}$) separated by a distance $d$. Then we could use either parametric or non-parametric methods to estimate $\rho_r(\cdot)$ based on \{$\tilde\rho_r(d): d\in\CD$\}.
Many classical parametric models in spatial statistics \citep{cressie2015statistics} can be used. 
Taking as an example, we perform a weighted least square approach \citep{sherman2011spatial} on \{$\tilde\rho_r(d): d\in\CD$\} to fit an exponential model $\exp(-|\bs_i-\bs_j|/\phi)$ with the scale parameter $\phi$.
Then $\rho_r(d)$ is estimated by $\exp(-d/\hat \phi_r)$ with the weighted least square estimator $\hat \phi_r$ for each $r$.
Alternatively, a smooth estimate of $\rho_r(\cdot)$ could also be obtained by a nonparametric regression approach, such as a local linear estimation \citep{fan1996local}  on \{$ d,\tilde\rho_r(d): d\in\CD$\}.
We evaluate the performance of parametric and non-parametric methods via simulation studies in Section \ref{simulation}.

An important issue is the choice of the spatial lag $h$. 
For spatial data, it is common that observations separated by smaller lags are more correlated. Evidence for any departure from weak separability is usually stronger when the test statistic is formed by using a smaller $h$, at least as compared to using a much larger lag at which the correlation may be negligible. For data observed on a regular space $\CS \subset \mathbb{Z}^2$, where $\mathbb{Z}^2$ can be regarded as the two-dimensional space of the integer lattice points with minimum grid distance, saying $d_0=1$, we could naturally use a lag-$z$ covariance estimator where $z$ is a positive integer such that $h=z d_0$. 
{Our simulation results in Section \ref{simulation} provide empirical support for using small lags, which also coincides with the theoretical consideration of small neighborhood of $h$.}

\begin{remark}\label{multiple test}
 \rm  We also consider combining information from a range of $h$ using multiple tests, and provide the relevant results based on the Bonferroni correction in Section S.3.2 of the Supplementary Material. It is seen that this testing procedure combining different lags has reasonable size, but is less powerful than the test using lag-1 covariance. This is expected given that our simulation results in Section 4 also indicate that the test using lag-1 covariance is the most powerful.
\end{remark}

	\section{Simulation Study}\label{simulation}

In this section we assess the performance of the proposed test. Set the time domain $T=[0,1]$ with 100 equally spaced time points, and the spatial region be a regular grid on $D=[0,2] \times [0,2]$ whose spatial grid increment is $0.05$, i.e., the number of spatial points $N=40^2=1600$. We generate $X(\bs,t)$ from the model,
$X(\bs,t) = \mu(t) + \sum_{r=1}^p \xi_r(\bs)\psi_r(t)$,
using a mean function $\mu(t)=3+2t^2$ as in \cite{li2014functional}, 
and $p=10$ basis functions, $\psi_{r}(t)=\sqrt{2}\cos ( r\pi t )$ when $r$ is odd and $\psi_{r}(t)=\sqrt{2}\sin\{(r-1)\pi t\}$ when $r$ is even as in \cite{kong2016partially}.
The \{$\xi_r(\cdot)$\} are generated marginally from Gaussian random fields, with isotropic Mat\'ern covariance structures $c_r(\bs_1-\bs_2;\nu,\phi) = \omega_r\mbox{M}( |\bs_1-\bs_2|;\nu,\phi)$ with $\omega_r=4r^{-2}$ and
$$\mbox{M}(d;\nu,\phi) = \frac{2^{1-\nu}}{\Gamma(\nu)}(\frac{d}{\phi})^{\nu}K_{\nu}(\frac{d}{\phi}).$$
In the above, $K_{\nu}$ is the modified Bessel function of the second type of order $\nu$, where $\nu$ controls the smoothness of the process, and
$\phi>0$ is the range parameter controlling the rate of decay of spatial correlation, where a larger $\phi$ corresponds to a stronger correlation \citep{cressie2015statistics}. We set $(\phi_1,\phi_2,\phi_3,\phi_4)=(0.2,0.1,0.15,0.08)$, and $\phi_5,\dots,\phi_{10}=0$ (i.e., no spatial correlation) for the high-order principal components.  

To study the power, we generate $\{ \xi_1(\cdot),\xi_2(\cdot) \} $ from a bivariate Mat\'ern model \citep{gneiting2010matern} using the above marginal covariance functions:
$$ c_{1}(d)=\omega_1 \mbox{M}(d;\nu_{1}=1,\phi_{1}=0.2), \quad c_{2}(d)=\omega_2 \mbox{M}(d;\nu_{2}=0.5,\phi_{2}=0.1), $$
and the following cross-covariance function:
$$c_{12}(d)=\rho_{12}\sqrt{\omega_1\omega_2} \mbox{M}(d;\nu_{12}=0.8,\phi_{12}=0.15).$$	
Here the coefficient $\rho_{12}=0$ yields the null hypothesis, i.e., the spatial random fields $\xi_r(\cdot)$ are mutually uncorrelated. The correlation between $\xi_1(\cdot)$ and $\xi_2(\cdot)$ increases as $\rho_{12}$ becomes larger, facilitating the departure from weak separability. The setup of $(\nu_1,\nu_2,\nu_{12})$ satisfies the condition in Theorem 3 of \cite{gneiting2010matern}, such that the above bivariate Mat\'ern model is valid. The other \{$\nu_{r}$\} are all set to be $0.5$ for $3 \le r \le 10$, noting that $\mbox{M}(d;\nu=0.5,\phi)$ reduces to the exponential model.

We implement the parametric and nonparametric methods in Section \ref{implementation} based on the empirical lag covariance with the 0.05 significance level. The empirical size and power are assessed with 1000 Monte Carlo runs. We investigate the test performance using different FVE threshold values (80\%, 90\% or 95\%), shown in Table \ref{rejection rate for grid data} the empirical rejection rate results using the lag-$z$ covariance defined in Section \ref{implementation} from $z=1$ to 4.  
We can see that when $\rho_{12}=0$, both the parametric and non-parametric tests have reasonable sizes across different lag choices, and  
the rejection rates rise rapidly for all the methods as $\rho_{12}$ grows from 0.2 to 0.6.
It is clear that the lag-1 covariance is the most powerful and the test performance deteriorates substantially as the lag increases, especially when $\rho_{12}=0.4$ or 0.6. 
This supports the advantage of using small lags, as discussed in Section \ref{implementation}, in order to detect departures from weak separability. 
Therefore, in practice we suggest to focus more on the test results based on the lag-1 covariance estimation for data collected on a regular spatial grid.

\begin{table}[h!]
      \caption{Rejection rates (\%) for the weak separability tests based on the lag covariance under different FVEs, with the asymptotic covariance estimated by parametric (Para) and non-parametric (Nonp) methods. 
 } 
      \centering
           \begin{tabular}{cccccccccccccccc}
                 \hline\hline
                & &\multicolumn{2}{c}{lag-1}& &\multicolumn{2}{c}{lag-2} & &\multicolumn{2}{c}{lag-3} & &\multicolumn{2}{c}{lag-4} \\
                \cline{3-4} \cline{6-7} \cline{9-10} \cline{12-13}
                $\rho_{12}$  & $\mbox{FVE} $ & Para &Nonp & & Para &Nonp && Para &Nonp  && Para &Nonp  \\ 
               \hline
               0& 80\%&  4.2 &  3.4 & &  6.4 & 4.2  && 7.4& 5.3 && 6.6 & 5.4  \\ 
               &90\% &   4.7 & 5.4   & & 6.4 &4.2  &&7.3 &5.2  && 6.5& 5.3  \\ 
               &95\% & 5.6  & 6.9 && 7.9 & 8.3 && 6.6 & 6.5  && 6.1& 5.2   \\ 
               \hline
               0.2& 80\% &    85.0 & 87.5 &&76.5 & 69.5  && 33.0 & 32.5 && 12.0& 15.5  \\ 
               &90\% &   74.5  & 79.5 &&74.0 & 67.5 & &32.0 & 32.0 && 12.0 & 15.5 \\ 
               &95\%& 70.5  & 77.5 &&63.5 & 62.5 & &27.5 & 31.5 && 12.0 & 15.5  \\ 
               \hline
               0.4 & 80\% &  100 & 100 && 94.0& 94.0 &&68.0 & 69.0 &&32.5 & 37.0   \\ 
               &90\%  &95.0  &94.5  &&95.0 &94.5 &&68.5& 69.5  &  &32.5 & 37.5   \\ 
               &95\%&100  &100  &&98.5 &98.0 &&73.5& 77.0  && 31.0 & 39.0   \\ 
               \hline
               0.6 & 80\% &  100& 88.0 &  & 89.0 & 82.0 && 67.5 & 68.0 && 37.0& 41.0  \\ 
               &90\%  &100 &100  &&90.0 &91.0 &&68.0& 68.0 && 37.0 & 41.0   \\ 
               &95\%&100 &100  &&99.5 &99.5 &&79.5& 81.5 && 41.5 & 47.5   \\ 
               \hline
               \hline
                 \end{tabular}
      \label{rejection rate for grid data}
\end{table}

Note that, for the lag-1 covariance, $R_N$ is selected to be 2 in nearly all 1000 trials when the FVE is set to 80\%. When FVE$\,=\,$90\%, the proportions of instances in which $R_N=2$ and $R_N=3$ are 33\% and 66\%, respectively, and when FVE$\,=\,$95\% the proportion is 91\% for $R_N=3$ and 9\% for $R_N=4$.
As explained in Section \ref{implementation}, the proposed test is usually stable under the null hypothesis for different $R_N$ values, though the size may be inflated for overly large $R_N$. 
On the other hand, the results under alternative depends on the occurrence of non-separable components. In our settings, the correlation emerges for the first two FPC scores, thus the power seems to be better if $R_N=2$ is mostly selected, i.e., FVE=80\%, which is seen in Table \ref{rejection rate for grid data}. As expected, the power increases as $\rho_{12}$ grows that makes the first two components more non-separable.
As a practical guidance, the claim of weak separability should be a comprehensive conclusion across different $R_N$ or FVE values. More discussion and simulation regarding the choice of $R_N$ are given in Section S.3.1 of the Supplementary Material.
Lastly we see that the results based on parametric and nonparametric modeling of correlation are very similar.
We also perform the tests with different range parameters (see Section S.4 of the Supplementary Material) and different space/time window (not reported for space concerns), and both studies yield similar results. 

	\section{Real Data Application}\label{real data}

	\subsection{China PM$_{2.5}$ data}\label{China's PM2.5 data}

Chronic and severe air pollution has affected a significant portion of China in recent years. 
Among all air pollutants, the fine particulate matter with aerodynamic diameters less than 2.5 $\mu m$, also known as PM$_{2.5}$, is considered to have the most damaging effect on health. 
It is a common practice to analyze time-varying environmental variables using functional data analysis, see examples of temperature and log-precipitation curves in \cite{RamsayJ.O.JamesO.2005Fda/} and wind speed data in \cite{constantinou2017testing}. 
Recent research on statistical modeling of China PM$_{2.5}$ has also received considerable attention, see \cite{LiangXuan2015ABP2} and \cite{zhang2017cautionary}. 


The Chinese government started monitoring PM$_{2.5}$ concentrations from 2013 and have established a large national monitoring network for air quality assessment by 2017. Real-time measurements of major pollutants across nearly 1,500 monitoring sites in 369 cities are continuously recorded and sent to the China National Environmental Monitoring Center (CNEMC) \citep{zhang2017cautionary, wu2018probabilistic}.
The PM$_{2.5}$ data used in our analysis are constructed based on the Nested Air Quality Prediction Modeling System (NAQPMS), which is a multi-scale chemical transport model proposed by the Institute of Atmospheric Physics, Chinese Academy of Sciences \citep{wang2006development}. The NAQPMS simulates the chemical and physical processes of air pollutants by solving the mass balance equation using terrain-following vertical coordinates. 
 The output PM$_{2.5}$ concentrations cover the entire China and have a $15$ km $\times$ $15$ km horizontal resolution on a regular $432\times339$ spatial grid. 

China has a topologically diverse landscape, and the entire spatial region
may be too large to impose stationarity, 
we choose 6 subregions according to the topographic division of China: the North China Plain (NCP), Yangtze River Delta (YRD), Pearl River Delta (PRD), Sichuan Basin (SB), Xinjiang (XJ) and Tibet (TB). 
Figure \ref{PM_map} displays an overview of these regions. 
In each region we extract a $40\times40$ grid from the whole dataset for ease of computation, with hourly recordings from December 1 to 30, 2016. 
Based on averaging every 4 consecutive hours, the number of time points is $30$ (days) $\times$ $6 =180$.
To check the spatial stationarity we perform a sensitivity analysis, and the results (provided in the Section S.1 of the Supplementary Material) suggest that there is no serious violation.

        \begin{figure}[h!]
                         \centering
        			\includegraphics[width=0.7\linewidth]{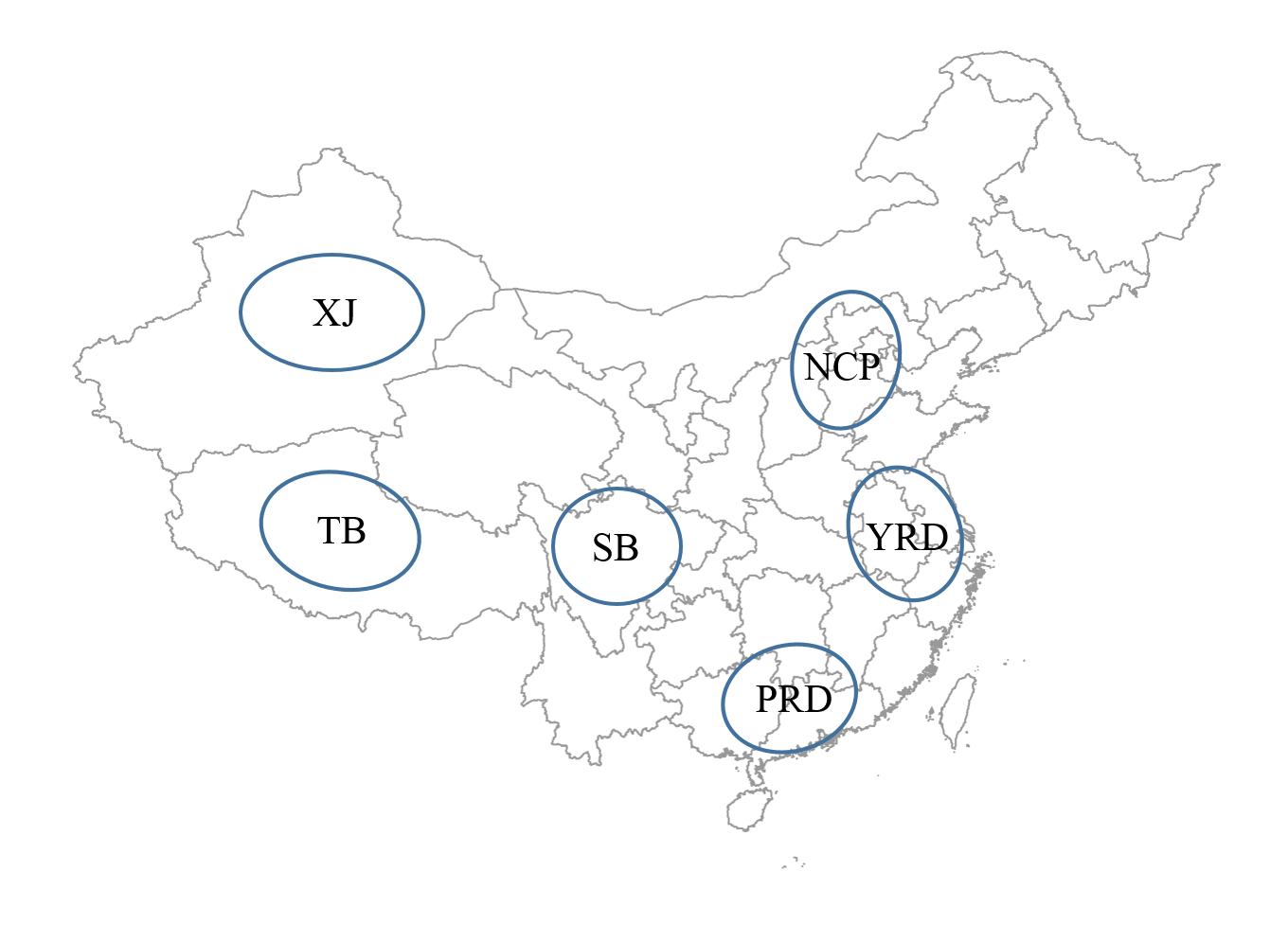}
        			\caption{Locations of the 6 subregions in the weak separability tests for China PM$_{2.5}$ data.}
        			\label{PM_map}
        	\end{figure}

We apply the proposed test in the above 6 regions based on the lag-1 covariance due to its advantage from methodological and empirical perspectives, with truncation of 80\%, 90\% and 95\% FVEs respectively. The resulting $p$-values are summarized in Table \ref{China's PM2.5 results}. We can see that the results of the parametric and nonparametric methods using different FVEs are in agreement: the hypothesis of weak separability is rejected for XJ and TB but not for the other 4 regions. 
This is an interesting phenomenon because the four non-rejected regions are relatively more developed areas in China. 
It has been reported that the formation and transmission mechanism of PM$_{2.5}$ exhibits completely different patterns in XJ and TB located mainly in the deserts and plateaus of western China \citep{Wang2015}.
Based on our analysis,
it is reasonable to assume weak separability for modeling the PM$_{2.5}$ data in east-central China, but not in XJ or TB regions.

\begin{table}[h!]
\centering
 \caption{The $p$-values of weak separability tests using parametric (Para) and nonparametric (Nonp) methods with different truncation levels for China PM$_{2.5}$ data.}
\begin{tabular}{ccccccc}
  \hline \hline
Region& NCP  & YRD &  PRD &  SB &  XJ & TB    \\
\hline
{FVE = 80\% }     & $R_N = 2$ & $R_N = 3$ & $R_N = 2$ & $R_N = 3$ & $R_N = 2$ & $R_N = 3$ \\
Para    &0.136 & 0.925 & 0.722 &0.937  &0.002&0.000  \\ 
 Nonp   &0.287 & 0.620 & 0.709 &0.663  &0.000&0.000  \\ 
  \hline
FVE  = 90\%     & $R_N = 3$ & $R_N = 6 $ & $R_N = 3$ & {$R_N = 6$} & {$R_N = 5$}& {$R_N = 5$}\\
Para   & {0.150} &  {0.863} & {0.713} & {0.535}  &{0.000}&{0.000}  \\ 
 Nonp  & {0.212} &  {0.399} & {0.690} & {0.440}  &{0.000}&{0.000}  \\ 
  \hline
 FVE = 95\%    & ${R_N = 6}$& {$R_N = 9$} & {$R_N = 6$} & {$R_N = 9$} & {$R_N = 8$} & {$R_N = 8$} \\
Para                    & {0.759}  & {0.997} & {0.521} & {0.366}  &{0.000}&{0.000}  \\ 
 Nonp                  &{0.289}  &{0.119} &{0.230} &{0.117}  &{0.000}&{0.000}  \\ 
  \hline \hline
\end{tabular}
\label{China's PM2.5 results}
\end{table}


%


	\subsection{Harvard Forest data}\label{harvard}
This dataset consists of the Enhanced Vegetation Index (EVI) series at Harvard Forest, which were previously studied by \cite{liu2017functional}. 
The EVI is calculated from surface spectral reflectance measurements collected from moderate-resolution imaging spectroradiometers onboard NASA's Terra and Aqua satellites. Specifically, the data are extracted for a 25 by 25 pixel window (covering an area of approximately 134 km$^2$), centered over the Harvard Forest Long Term Experimental Research site in Petershan, MA. 
The EVI data are recorded from 2001 to 2006 at 8-day intervals, see \cite{liu2017functional} for more details. 
By averaging 3 consecutive temporal observations to reduce noise, the dataset used in our analysis has 625 spatial grids and 92 time points.
		
\cite{liu2017functional} proposed a spatial PACE model based on \cite{YaoFang2005FDAf} to reconstruct the spatial functional data and perform an isotropy test. 
According to the study of \cite{liu2017functional}, it is reasonable to assume spatial stationarity, which is also verified by our sensitivity analysis similar to that used for the China's PM$_{2.5}$ data example.
We notice that weak separability is assumed in their spatial functional model and forms a foundation for subsequent analysis. 
We now examine the appropriateness of weak separability, and apply the proposed test based on the lag-1 covariance. The results show that the FVEs for the first two and three components are respectively 69.8\% and 78.5\%. The corresponding $p$-values using the parametric and nonparametric methods are 0.872 and 0.880, respectively, when $R_N$ is 2, and are 0.073 and 0.079 when $R_N$ is 3. Interestingly, as $R_N$ increases to 4 (corresponding to FVE=84.1\%), the $p$-values become 0.018 and 0.019, then they decrease to less than 0.01 when more than 5 components are considered. 
This indicates that the correlated FPC fields that violate the weak separability assumption do not emerge until $R_N=4$. 
Our results suggest that, 
if considering a rough representation with no more than 3 components explaining around 80\% FVE, the model used in \cite{liu2017functional} which focuses primarily on the first two FPCs seems reasonable. However, if we are interested in a comprehensive spatio-temporal functional data model that contains more components, the weak separability assumption appears not valid.

	\section{Discussion and Conclusion}\label{discussion}

In this work we introduce a sensible definition of weak separability for spatial functional field. 
This flexible yet parsimonious representation views the space and time domains from different perspectives with ease of modeling/computation and interpretation. 
By means of the lag covariance function, we develop a formal hypothesis test based on the asymptotic distribution that is easy to handle under Gaussian assumption and does not require computationally intensive resampling procedures, such as bootstrap approximations. 
In particular, our methods are motivated by (and more applicable to) the typical non-replicated spatio-temporal data. 
We implemented the test procedures for the Harvard Forest data and China PM$_{2.5}$ data, and obtained interesting and insightful results. 
We recommend to exploit the proposed test prior to further statistical analysis that often imposes weak separability assumption for feasible modeling.  

For future work, the proposed procedure deals with data observed on a dense temporal grid, which can be extended to the case of sparse functional observations with measurement errors \citep{YaoFang2005FDAf, liu2017functional} through an adaptation of lag covariance estimation. For instance, 
if the data are spatially irregularly spaced, one could implement the kernel smoothing methods mentioned in Section \ref{concept of covariance estimators} with more involved technical development.
For those whose data show obvious spatial anisotropy, the lag covariance in some specific directional lag can be used. Methods for testing for isotropy, such as those in \cite{guan2004nonparametric} and \cite{liu2017functional}, may also be studied in current context.
The discussion and results in Section S.2 of the Supplementary Material provide some evidence for the sensitivity of our methods to the Gaussian assumption, and demonstrate the favorable performance of our method over other numerical methods such as the block bootstrap. 
However, more studies about 
the approximating method on the asymptotic covariance without the Gaussian assumption are still called for and exhibit challenges for the non-replicated spatio-temporal data.
Another potential topic is spatio-temporal point process, where the spatial functional process act as a latent effect and is related to the intensity function of the point process with a nonlinear link function \citep{li2014functional}. 
	
\section*{Supplementary Material}

For space economy, we collect more discussions and additional results on spatial stationarity and Gaussian assumption, some implementation issues for the choices of truncation parameter and spatial lag, and some technical proofs of the propositions and lemmas in an online Supplementary Material.

%
%
%


\begin{appendix}	

		\Appendix	
\section{Notation}\label{A1}
We first introduce some notations in Hilbert spaces. 
Let $\CH$ be a real separable Hilbert space with the usual inner product $\left< f,g \right>=\int f(t)g(t)dt$. 
As standard definitions in \cite{aston2017tests} 
the space of Hilbert-Schmidt operators on $\CH$ is denoted as $\CB_{HS}(\CH)$, and is a Hilbert space with the inner product $\left<K_1,K_2\right>_{HS}=\sum_{i\geq 1}\left<K_1(e_i),K_2(e_i)\right>$ and the induced norm $\|\cdot\|_{HS}$. 
For $z,y \in \CH$, $z\otimes y$ is the operator defined by $(z\otimes y)(x) = \left<z,x\right>y$, and so for $z,y \in \CB_{HS}(\CH)$.
A spatial functional field $X(\bs,t)$ at fixed $\bs$ is denoted as $X(\bs) \in L^2(\CT)$ with the mean function $\mu(\bs) =\E X(\bs)$.
The covariance operator of $X(\bs)$ is defined by 
$ \BC_{\bs,\bs} = \E \left[ \left\{ X(\bs)-\mu(\bs) \right\} \otimes \left\{ X(\bs)-\mu(\bs) \right\} \right] $, and
the cross covariance operator at fix locations $\bs_1$ and $\bs_2$ is defined as
$ \BC_{\bs_1,\bs_2} = \E \left[ \left\{ X(\bs_1)-\mu(\bs_1) \right\} \otimes \left\{ X(\bs_2)-\mu(\bs_2) \right\} \right] $ \citep{Hsing2015Theoretical, hormann2011consistency}.
For a spatially stationary functional field $X(\bs,t)$ with the mean function as $\mu =\mu (\bs)$, the covariance estimator 
\begin{equation*}\label{covariance operator}
\BC = \E \left[ \left\{ X(\bs)-\mu \right\} \otimes \left\{ X(\bs)-\mu \right\} \right],
\end{equation*}
and the lag covariance operator $\BC^{(h)}$ is defined by 
\begin{equation*}\label{lag covariance operator}
\BC^{(h)} = \E \left[  \left\{ X(\bs_1)-\mu  \right\} \otimes \left\{ X(\bs_{2})-\mu \right\} \right],  \mbox{ where } |\bs_1-\bs_2|=h,
\end{equation*}
under isotropy.
The covariance function $C(t_1,t_2)$ and $C^{(h)}(t_1,t_2)$ defined in Section \ref{concept of covariance estimators} can be seen as the kernel of the operator $\BC$ and $\BC^{(h)}$; see Chapter 7.2 and 7.3 of \cite{Hsing2015Theoretical} for more details about the (cross) covariance operators in Hilbert space. 
Assume we observe $\{X(\bs_i)\}_{i=1,\dots,N}$ from $X(\bs)$, and denote $X_i=X(\bs_i)$. 
We estimate $\mu$ by the sample mean $\bar{X}=N^{-1}\sum_{i=1}^NX_i$, and $\BC$ by the sample covariance operator 
\begin{equation}\label{empirical covariance operator}
\wh\BC = \frac{1}{N} \sum_{i=1}^N ( X_i-\bar{X} ) \otimes  (X_i-\bar{X} ).
\end{equation}
Let $\mathscr{A}_h = \{(i,i'): | \bs_i-\bs_{i'} | = h \}$ be the set of location pairs at lag $h$, and $N_h= |\mathscr{A}_h|$ be the cardinality of $\mathscr{A}_h$.
We then have the empirical lag covariance operator given by
\begin{equation*}\label{empirical lag covariance operator}
\wh\BC^{(h)}= \frac{1}{N_h} \sum_{(i,{i'})\in \mathscr{A}_h} ( X_i-\bar{X} )  \otimes (X_{i'}-\bar{X}) .
\end{equation*}
It can be seen that all these operators $\BC$, $\BC^{(h)}$, $\wh\BC$ and $\wh\BC^{(h)}$ belong to $\CB_{HS}\{ L^2(\CT) \}$ \citep{Hsing2015Theoretical}. 
To be more concise and without confusion, 
in what follows we denote 
$\psi(t)$ as $\psi$, $\hat\psi_r^{(h)}(t)$ as $\hat\psi_r^{(h)}$,  $\int C(t_1,t_2)\psi(t_1)dt_1$ as $\int \BC\psi$, and $\iint C(t_1,t_2)\psi_j(t_1)\psi_k(t_2)dt_1dt_2$ as $\int \BC\psi_j\psi_k$. 

\section{Proof of Theorem \ref{asymptotic express of Tn} }\label{A3}
{
We first state some technical lemmas for the main theorems. 
Lemmas \ref{asymptotic normality of iid covariance operators} and \ref{asymptotic normality of covariance operators} provide the asymptotic results for the covariance and lag covariance estimators.
Lemma \ref{moment bound} provides the perturbation results for the estimates obtained by FPCA, which serve as building blocks for establishing the moment bound result in Theorem \ref{asymptotic express of Tn}.
}

\begin{lm}\label{asymptotic normality of iid covariance operators}
Consider the covariance operator $\widehat\BC$ defined in (\ref{empirical covariance operator}) with $X_i \overset{\text{i.i.d}}{\sim} X \in L^2(\CT)$. If $\E \|X_i\|^4<\infty$, then $N^{1/2} (\wh \BC - \BC)$ converges in distribution to a mean-zero Gaussian random element in $\CB_{HS}\{L^2(\CT)\}$.
\end{lm}

\begin{remark}\label{clt}
 \rm Here $\E \|X_i\|^4<\infty$ is a sufficient condition for the CLT of covariance operators under the Hilbert-Schmidt topology. This is a weaker assumption than Condition 2.1 in \cite{aston2017tests} that 
 $\sum_{r=1}^\infty(\E ([ \left<X,e_r\right>^4 ] )^{1/4}<\infty$ for some orthonormal basis $(e_r)_{r\geq1}$, which is required to prove the weak convergence under the trace-norm topology \citep[see Remark 3.2 of][]{bagchi2020test}.
  \end{remark}

\begin{lm}\label{asymptotic normality of covariance operators}
Consider a spatially stationary functional field $X(\bs) \in L^2(\CT)$ with covariance and lag covariance operators defined in \ref{A1}.
Under Conditions \ref{condition: high order moment} and \ref{condition: mixing}, $[ N^{1/2} (\wh \BC - \BC), N_h^{1/2} \{ \wh \BC^{(h)} - \BC^{(h)} \}  ] $ converges in distribution to a mean-zero Gaussian random element in $\CB_{HS}\{L^2(\CT)\} \times \CB_{HS}\{ L^2(\CT)\} $.
\end{lm}

{
Define the set of realizations such that, for sample size $N$, some C and any $\tau<1$,
\begin{equation}\label{high probability set}
\CE_{N,R_N}= \left\{ \left( \hat\eta_j^{(h)} - \eta_k^{(h)} \right)^{-2} \leq 2 \left( \eta_j^{(h)} - \eta_k^{(h)} \right)^{-2} \leq C N^\tau, j,k = 1,\dots,R_N, j\not=k \right\}.
\end{equation}
\begin{lm}\label{moment bound}
(a) Under Conditions  \ref{condition: high order moment} to \ref{condition: truncation}, we have $\BP(\CE_{N,R_N}) \rightarrow 1$ as $N \rightarrow \infty$. \\
(b) Under Conditions  \ref{condition: high order moment} to \ref{condition: truncation}, for each integer $b>0$, on the high probability set $\CE_{N,R_N}$,
$$ \E \| \hat\psi_j^{(h)} - \psi_j \|^{2b} = O\{(j^2 N^{-1})^b\}, $$ and $O(\cdot)$ is uniform in $j=1,\dots,R_N$. \\
(c) Under Conditions  \ref{condition: high order moment} to \ref{condition: truncation}, on the high probability set $\CE_{N,R_N}$,
\begin{equation}\label{expansion of lag FPC}
 \hat\psi_j^{(h)}(t) -\psi_j(t)=\sum_{k:k\not=j}(\eta_j^{(h)}-\eta_k^{(h)} )^{-1}\psi_k(t) \int\left(\wh\BC^{(h)}-\BC^{(h)}\right)\psi_j\psi_k + \alpha_{j}(t),
 \end{equation}
where  $$\E \|\alpha_{j}\|^2= O(j^{2a+4}N^{-2})$$ uniformly in $j=1,\dots,R_N$. 
\end{lm}
}
{
\begin{remark}\label{bound}
 \rm  The moment bound for $ \| \hat\psi_j^{(h)} - \psi_j \|^{2b}$ with diverging truncation provides a stronger result  than the consistency of $\hat\psi_j^{(h)}$, i.e., $O_p(\cdot)$ type. The expansion (\ref{expansion of lag FPC}) follows directly from Lemma 1 of \cite{kong2016partially} or (5.22) of \cite{hall2009theory}.  
It has been shown that $ \|\alpha_{j}\| = O_p(j^{a+2}N^{-1})$ in \cite{kong2016partially}. 
However, to derive the convergence rate for $\|T_h(\mathbb{I}_{R_N})- T_h^*(\mathbb{I}_{R_N})\|^2$ in Theorem \ref{joint asymptotic normality of Tn}, which is based on a sum over a diverging truncation set of indices for the error terms, we need the moment bounds for $\|\alpha_{j}\|^2$ over this truncation set.
The proof of Lemma \ref{moment bound} is based on similar arguments and moment calculations in \cite{kong2016partially} and \cite{hall2009theory}; see Section S.5.2 of the Supplementary Material.
\end{remark}
}

 \begin{proof}[Proof of Theorem \ref{asymptotic express of Tn}]
Using the notations in \ref{A1}, we denote the eigen-decomposition in (\ref{eigen-decomposition lag}) as $\int \wh\BC^{(h)} \hat\psi_{r}^{(h)}=\hat\eta_{r}^{(h)}\hat\psi_{r}^{(h)}$, and similarly $\int\BC^{(h)} \psi_r=\eta_r^{(h)}\psi_r$.
Recall that 
$\hat\xi_{ir}^{(h)}=\int (X_i-\bar X)\hat\psi_{r}^{(h)}$, $i=1,\dots,N$.
Then $T_h(j,k)$ can be expressed by 
\begin{align*}
 \frac{1}{\sqrt{N}}\sum_{i=1}^N \hat\xi^{(h)}_{ij}\hat\xi^{(h)}_{ik} &=\frac{1}{\sqrt{N}}\iint\sum_{i=1}^N \left\{ X_i(t_1)-\bar X(t_1)\right\} \hat\psi_j^{(h)}(t_1) \left\{ X_i(t_2)-\bar X(t_2)\right\} \hat\psi_k^{(h)}(t_2) dt_1dt_2 \\
             &=\sqrt{N}\int\wh\BC\hat\psi_j^{(h)}\hat\psi_k^{(h)}=
             \sqrt{N}\int (\wh\BC-\BC)\hat\psi_j^{(h)}\hat\psi_k^{(h)} + \sqrt{N}\int \BC \hat\psi_j^{(h)} \hat\psi_k^{(h)}.
  \end{align*}
We shall show that both of the above two terms converge, and the leading terms are respectively $\sqrt{N}\int(\wh\BC-\BC)\psi_j\psi_k$ and $-{\sqrt{N}}/{\rho_{jk}^{(h)}} \int (\wh{\BC}^{(h)} -\BC^{(h)} )\psi_j\psi_k$. 
First, let 
$\Delta_1(j,k)  := \sqrt{N}\int (\wh\BC-\BC)\hat\psi_j^{(h)}\hat\psi_k^{(h)} -   \sqrt{N}\int(\wh\BC-\BC)\psi_j\psi_k$, then $\Delta_1(j,k) = \sqrt{N} \int (\wh\BC-\BC) (\hat\psi_j^{(h)}-\psi_j)\hat\psi_k^{(h)} + \sqrt{N} \int (\wh\BC-\BC) (\hat\psi_k^{(h)}-\psi_k) \psi_j$.
As $\|\psi_j\| = \|\hat\psi_k^{(h)}\| = 1$, we have $ | \Delta_1(j,k)  |  \leq  \sqrt{N} \,  \| \wh\BC-\BC\|_{HS} \,  \| \hat\psi_j^{(h)}-\psi_j \|  + \sqrt{N} \,  \| \wh\BC-\BC\|_{HS} \,  \| \hat\psi_k^{(h)}-\psi_k \|  \leq   \sqrt{N} \,  \| \wh\BC-\BC\|_{HS} \, ( \| \hat\psi_j^{(h)}-\psi_j \| + \| \hat\psi_k^{(h)}-\psi_k \|)   $. 
It then follows by Cauchy-Schwarz inequality that
$ (\E | \Delta_1(j,k)  |)^2  \leq  C_1 N \, \E \| \wh\BC-\BC\|_{HS}^2 \, (\E \| \hat\psi_j^{(h)}-\psi_j \|^2  + \E \| \hat\psi_k^{(h)}-\psi_k \|^2)  $, and
$ \E | \Delta_1(j,k)  |^2 \leq  C_2 N \{ \E \| \wh\BC-\BC\|_{HS}^4 \, (\E \| \hat\psi_j^{(h)}-\psi_j \|^4  + \E \| \hat\psi_k^{(h)}-\psi_k \|^4) \}^{1/2}  $ for some constants $C_1$ and $C_2$.
Note that $\E \| \hat\psi_j^{(h)}-\psi_j \|^4 = O(j^4N^{-2})$ uniformly in $j=1,\dots,R_N$ by Lemma \ref{moment bound}(b) with $b=2$, and $\E \| \wh\BC-\BC\|_{HS}^4 = O(N^{-2})$ according to the proof of Lemma  \ref{moment bound}(a).
It then follows that $\E |\Delta_1(j,k) |^2 = O( R_N^2 N^{-1} )$ uniformly in $j,k=1,\dots,R_N$ and $j\not=k$.

Next we will show the moment bound for $ \Delta_2(j,k) =  \sqrt{N}\int \BC \hat\psi_j^{(h)} \hat\psi_k^{(h)} + {\sqrt{N}}/{\rho_{jk}^{(h)}} \int (\wh{\BC}^{(h)} -\BC^{(h)} )\psi_j\psi_k$.
Denote $\bm{M}_{k,j} = (\eta_j^{(h)}-\eta_k^{(h)})^{-1}\int(\wh\BC^{(h)}-\BC^{(h)})\psi_j\psi_k$, the equation (\ref{expansion of lag FPC}) translates into
$ \hat\psi_j^{(h)}(t)-\psi_j(t)=\sum_{k:k\not=j}\bm{M}_{k,j}\psi_k(t)+ \alpha_{j}(t)$.
Then we have 
$$\sqrt{N}\int \BC \hat\psi_j^{(h)} \hat\psi_k^{(h)} =\sqrt{N}\int \BC \left(\psi_j+\sum_{r:r\not=j}\bm{M}_{r,j}\psi_r + \alpha_{j} \right) \left(\psi_k+\sum_{r:r\not=k}\bm{M}_{r,k}\psi_r + \alpha_{k} \right).$$
Note that $\int\BC\psi_j\psi_k=0$ for $j\not=k$ and $\int\BC\psi_j\psi_j=\omega_j$, it follows that
\begin{align*}
\sqrt{N}\int \BC \hat\psi_j^{(h)} \hat\psi_k^{(h)} &=\sqrt{N}\int \BC\psi_j(\bm{M}_{j,k}\psi_j) + \sqrt{N}\int\BC\psi_k\left(\bm{M}_{k,j}\psi_k \right)+\beta_{jk} \\
     &=\sqrt{N}\omega_j\bm{M}_{j,k} + \sqrt{N}\omega_k\bm{M}_{k,j} + \beta_{jk} \\
     &=\sqrt{N}(\omega_j-\omega_k)\bm{M}_{j,k}+ \beta_{jk} \\
     &=\sqrt{N}(\omega_j-\omega_k)(\eta_k^{(h)}-\eta_j^{(h)})^{-1}\int\left(\wh\BC^{(h)}-\BC^{(h)}\right)\psi_j\psi_k + \beta_{jk}.
\end{align*} 
{Here we denote $\beta_{jk}=\beta_1+\beta_2+\beta_3$, where $\beta_1 = \sqrt{N}\int \BC \hat\psi_j^{(h)} \alpha_k$, $\beta_2 =\sqrt{N}\int \BC  ( \sum_{r:r\not=j}\bm{M}_{r,j}\psi_r + \alpha_{j} )( \sum_{r:r\not=k}\bm{M}_{r,k}\psi_r )$ and $\beta_3 = \sqrt{N}\int \BC \alpha_j\psi_k$. 
Note that $ |\beta_1|^2 \leq N \|C\|_{HS}^2 \| \hat\psi_j^{(h)}\|^2 \|\alpha_k\|^2 $ and $\| \hat\psi_j^{(h)}\|=1$, it follows that $\E |\beta_1|^2 \leq N |C\|_{HS}^2 \E \|\alpha_k\|^2= O(k^{2a+4}N^{-1})$ by Lemma \ref{moment bound}, and similarly $\E |\beta_3|^2 = O(j^{2a+4}N^{-1})$. 
By Parseval's identidy, $ \|\sum_{r:r\not=j}\bm{M}_{r,j}\psi_r(t) \|^2=    \sum_{r:r\not=j} (\eta_j^{(h)} - \eta_r^{(h)})^{-2} \{ \int (\wh\BC^{(h)}-\BC^{(h)} )\psi_j\psi_r  \}^2 = \sum_{r:r\not=j} \bm{M}_{r,j}^2$, 
Then we have $ |\beta_2|^2 \leq N  \|C\|_{HS}^2 \|\hat\psi_j^{(h)}-\psi_j \|^2  ( \sum_{r:r\not=k}\bm{M}_{r,k}^2$), and it follows that $\E |\beta_2|^2 \leq N  \|C\|_{HS}^2 \{ \E  \|\hat\psi_j^{(h)}-\psi_j \| ^4 \, \E  (\sum_{r:r\not=k}\bm{M}_{r,k}^2 )^2 \}^{1/2} = O(j^2k^2 N^{-1})$ based on Lemma \ref{moment bound}(b) and equation (S.2) in the proof of Lemma  \ref{moment bound}. 
Combining these results we obtain $\E | \beta_{jk} |^2 = O(  j^2k^2N^{-1} + j^{2a+4}N^{-1} +  k^{2a+4}N^{-1})$, leading to $\E | \beta_{jk} |^2 = O(R_N^{2a+4}  N^{-1})$ uniformly in $j,k=1,\dots,R_N$ and $j\not=k$. 
}
As written, $\rho_{jk}^{(h)}=(\eta_j^{(h)}-\eta_k^{(h)})(\omega_j-\omega_k)^{-1}$, then $\sqrt{N}\int \BC \hat\psi_j^{(h)} \hat\psi_k^{(h)} = -\sqrt{N}/\rho_{jk}^{(h)}\int\left(\wh\BC^{(h)}-\BC^{(h)}\right)\psi_j\psi_k + \Delta_2(j,k)$, and $\E |\Delta_2(j,k)|^2 = O(R_N^{2a+4}  N^{-1})$ uniformly in $j,k=1,\dots,R_N$ and $j\not=k$.
 
Now we denote
\begin{equation}\label{Th'}
T_h'(j,k)=\sqrt{N}\int \left( \wh\BC-\BC \right) \psi_j\psi_k-\sqrt{N}/\rho_{jk}^{(h)} \int\left(\wh\BC^{(h)}-\BC^{(h)} \right)\psi_j\psi_k.
\end{equation}
Note that $\int\BC\psi_j\psi_k=\int\BC^{(h)}\psi_j\psi_k=0$ under weak separability, then
\begin{align*}
T'_h(j,k)=&\sqrt{N} \int ( \wh\BC-\frac{1}{\rho_{jk}^{(h)}}~ \wh\BC^{(h)} ) \psi_j\psi_k \\
            =&\frac{1}{\sqrt{N}}\int \{ \sum_{i=1}^N (X_i-\bar{X})(X_i-\bar{X}) \} \psi_j\psi_k  - \frac{\sqrt{N}}{\rho_{jk}^{(h)} N_h} \int \{\sum_{(i,{i'})\in \mathscr{A}_h} (X_i-\bar X)(X_{i'}-\bar X) \} \psi_j\psi_k 
\end{align*} 
By the definition of $T_h^*(j,k)$ in (\ref{true term}), 
\begin{align*}
T^*_h(j,k)   =&\frac{1}{\sqrt{N}}\sum_{i=1}^N \xi_{ij}\xi_{ik} - \frac{\sqrt{N}}{\rho_{jk}^{(h)} N_h}\sum_{(i,i')\in\mathscr{A}_h} \xi_{ij}\xi_{i'k} \\ 
        =&\frac{1}{\sqrt{N}}\int \{ \sum_{i=1}^N (X_i-\mu)(X_i-\mu) \} \psi_j\psi_k - \frac{\sqrt{N}}{\rho_{jk}^{(h)} N_h} \int \{\sum_{(i,{i'})\in \mathscr{A}_h} (X_i-\mu)(X_{i'}-\mu) \} \psi_j\psi_k 
\end{align*} 
{Note that $N^{-1} \sum_{i=1}^N (X_i-\mu)\otimes(X_i-\mu)  - N^{-1}\sum_{i=1}^N (X_i-\bar{X})\otimes(X_i-\bar{X}) = (\bar{X}-\mu) \otimes (\bar{X}-\mu) $, 
$N_h^{-1} \sum_{(i,{i'})\in \mathscr{A}_h} (X_i-\mu)\otimes(X_{i'}-\mu) -N_h^{-1}\sum_{(i,{i'})\in \mathscr{A}_h} (X_i-\bar{X})\otimes(X_{i'}-\bar{X}) = (\bar{X}-\mu) \otimes (\bar{X}-\mu) $, 
and $\E \, \|(\bar{X}-\mu) \otimes (\bar{X}-\mu) \|_{HS} = O(N^{-1})$ by Proposition 2 of \cite{hormann2011consistency}.
This leads to that $\E |T'_h(j,k) - T^*_h(j,k) |^2 = O(N^{-1})$.
{
Combining the moment bound results of $\Delta_1(j,k)$ and $\Delta_2(j,k)$, we thus have $\E |T_h(j,k) - T^*_h(j,k) |^2 = O(R_N^{2a+4}  N^{-1})$ uniformly in $j,k=1,\dots,R_N$, for $j\not=k$.
It then follows by Condition \ref{condition: truncation} that $\E |T_h(j,k) - T^*_h(j,k) |^2 = o(1)$ and $|T_h(j,k) - T^*_h(j,k) | = o_p(1)$ by Chebyshev's inequality, where both $o(\cdot)$ and $o_p(\cdot)$ are uniform in $j,k$.
}
}
            
 \end{proof}


\section{Proof of Theorem \ref{joint asymptotic normality of Tn} and Corollary \ref{asymptotic under gaussian}    }\label{proof2}
 \begin{proof}[Proof of Theorem \ref{joint asymptotic normality of Tn} ]
{ 
We first show that $ \| T_h(\mathbb{I}_{R_N}) - T_h^*(\mathbb{I}_{R_N}) \| = o_p(1)$ on $\CE_{N,R_N}$, where $ \| T_h(\mathbb{I}_{R_N}) - T_h^*(\mathbb{I}_{R_N}) \|^2 = \sum_{(j,k)\in\mathbb{I}_{R_N}}  | T_h(j,k) - T_h^*(j,k) |^2$. 
In fact, $\E  \| T_h(\mathbb{I}_{R_N}) - T_h^*(\mathbb{I}_{R_N}) \|^2 = \sum_{(j,k)\in\mathbb{I}_{R_N}} \E | T_h(j,k) - T_h^*(j,k) |^2 = O(R_N^* R_N^{2a+4} N^{-1}) = O(R_N^{2a+6} N^{-1})$ by Theorem \ref{asymptotic express of Tn}, which leads to $\| T_h(\mathbb{I}_{R_N}) - T_h^*(\mathbb{I}_{R_N}) \|^2 = o_p(1)$ using Chebyshev's inequality and Condition {\ref{condition: truncation}$^*$}.
}
In addition, denote the vector by stacking $\{T_h’(j, k): (j, k)\in \mathbb{I}_{R_N}\}$ as $T_h'(\mathbb{I}_{R_N})$, where $T_h’(j, k)$ is defined in (\ref{Th'}), we could also obtain that
$\| T_h(\mathbb{I}_{R_N}) - T_h'(\mathbb{I}_{R_N}) \| = o_p(1)$ by noting $\E \|T_h(j,k) - T_h'(j,k) \|^2 = O(R_N^{2a+4} N^{-1})$ uniformly in $j, k$ in the proof of Theorem \ref{asymptotic express of Tn}.
 
According to Lemma \ref{asymptotic normality of covariance operators} and the continuous mapping theorem in metric space \citep{van2000asymptotic}, the sequence $\BZ_N = \sqrt{N} \{ ( \wh\BC-\BC) -(\wh\BC^{(h)}-\BC^{(h)}) /\rho_{jk}^{(h)} \}$ converges in distribution to a mean-zero Gaussian random element $\BZ$ in $\CB_{HS}\{L^2(\CT)\}$ under $N_h \asymp N$,
i.e., $ \pi (\BZ_N, \BZ) \rightarrow 0 $ due to the equivalence of weak convergence and Prokhorov metric \citep[][Section 6 of Chapter 1]{billingsley1999convergence}.
Then consider the continuous mapping $\tau_1$ from $\CB_{HS}\{L^2(\CT)\}$ to $\mathbb{R}^{R_N^*}$ that $\tau_1(\BX)_{j,k} = \int \BX \, \psi_j \psi_k$, $(j,k) \in \mathbb{I}_{R_N} $, that is, for any $\BX\in \CB_{HS}\{L^2(\CT)\}$, $\tau(\BX)$ is a $R_N^*$-dimensional vector with the $(j,k)$th element being $ \int \BX \, \psi_j \psi_k$.
Denote the multivariate normal vector \{$ \int \BZ \, \psi_j \psi_k: (j,k) \in \mathbb{I}_{R_N} $\} as $\CZ_{R_N^*}$, we thus have 
$$\pi \left( T_h'(\mathbb{I}_{R_N}),  \CZ_{R_N^*} \right) \le \pi \left(\BZ_N, \BZ \right) $$
using the mapping Theorem 3.2 of \cite{whitt1974preservation} and $\sup_{\BX\not=\BY} \frac{\| \tau(\BX)-\tau(\BY) \| }{ \| \BX-\BY\|_{HS} } \leq 1$ for $\BX,\BY \in \CB_{HS}\{L^2(\CT)\}$, where $\|\cdot\|$ is the Euclidean norm in $\mathbb{R}^{R_N^*}$, and 
for an operator $\BX \in \CB_{HS}\{L^2(\CT)\}$,  $\| \BX \|_{HS} = \sum_{r\geq 1} \left< \BX\psi_r, \BX\psi_r\right>$ by the definition in \ref{A1}. 
In fact, any operator $\BX,\BY \in \CB_{HS}\{L^2(\CT)\}$ can be expressed with $\BX = \sum_{j,k=1}^\infty a_{jk}\psi_j \otimes \psi_k$ and  $\BY = \sum_{j,k=1}^\infty \tilde{a}_{jk}\psi_j \otimes \psi_k$, where $a_{jk} = \left<\BX\psi_j,\psi_k\right> = \tau(\BX)_{j,k}$ and $\tilde a_{jk} = \left< \,\BY\psi_j,\psi_k \right> = \tau(\,\BY)_{j,k}$.
Following the definition of $\| \cdot \|_{HS} $, we have $\| \BX - \BY \|_{HS}^2= \sum_{j,k=1}^\infty (a_{jk}-\tilde a_{jk})^2$, which is not less than $ \| \tau(\BX) - \tau(\,\BY) \|^2 =\sum_{j,k\in \mathbb{I}_{R_N}} ( a_{jk}-\tilde a_{jk})^2 $.
As $ \| T_h(\mathbb{I}_{R_N}) - T_h'(\mathbb{I}_{R_N}) \| \xrightarrow{p} 0$, we then have $\pi \left( T_h(\mathbb{I}_{R_N}), \CZ_{R_N^*} \right) \rightarrow 0$.

To derive the covariance matrix $\Gamma$, we introduce the explicit expression of $\mathscr{A}_h= \{(i_1,i'_1),(i_2,i'_2),\dots,(i_{N_h},i'_{N_h}) \}$, the vectors $\bm\xi_r=(\xi_{1r},\dots,\xi_{Nr})^{\trans}$, $\bm\xi_{r,1}=(\xi_{i_1r},\dots,\xi_{i_{N_h}r})^{\trans}$,  $\bm\xi_{r,2}=(\xi_{i'_1r},\dots,\xi_{i'_{N_h}r})^{\trans}$. 
Then we obtain
\begin{equation}\label{asymptotic distribution, vector} 
T_h^*(j,k) =  \frac{1}{\sqrt{N}} \bm\xi_j^{\trans}\bm\xi_k - \frac{\sqrt{N}}{\rho_{jk}^{(h)} N_h} \bm\xi_{j,1}^{\trans}\bm\xi_{k,2},
\end{equation}
As a consequence, each element $\Gamma_{(j,k),(j',k')}$ in the covariance matrix $\Gamma$ of $T_h^*(\mathbb{I}_{R_N})  $
can be calculated by
\begin{equation}\label{asym covariance}
  \E \left( \frac{1}{\sqrt{N}} \bm\xi_{j}^{\trans}\bm\xi_{k} - \frac{\sqrt{N}}{\rho_{jk}^{(h)} N_h} \bm\xi_{j,1}^{\trans}\bm\xi_{k,2} \right) 
\left( \frac{1}{\sqrt{N}} \bm\xi_{j'}^{\trans}\bm\xi_{k'} - \frac{\sqrt{N}}{\rho_{j'k'}^{(h)} N_h} \bm\xi_{j',1}^{\trans}\bm\xi_{k',2} \right), 
\end{equation} 
which depends on the fourth moments $ \E ( \bm\xi_{j}^{\trans}\bm\xi_{k} \bm\xi_{j'}^{\trans}\bm\xi_{k'} )$, $\E ( \bm\xi_{j,1}^{\trans}\bm\xi_{k,2} \bm\xi_{j'}^{\trans}\bm\xi_{k'} ) $, $\E ( \bm\xi_{j}^{\trans}\bm\xi_{k} \bm\xi_{j',1}^{\trans}\bm\xi_{k',2} ) $ and $\E ( \bm\xi_{j,1}^{\trans}\bm\xi_{k,2} \bm\xi_{j',1}^{\trans}\bm\xi_{k',2} ) $. 
\end{proof}

 \begin{proof}[Proof of Corollary \ref{asymptotic under gaussian} ]
The assumption that $X(\bs,t)$ is Gaussian in Corollary \ref{asymptotic under gaussian} implies that the random fields $\xi_r(\bs) = \int (X(\bs) - \mu) \psi_r $ are jointly Gaussian, and jointly independent due to the weak separability. 
Denote the first term of (\ref{asymptotic distribution, vector}) as $J_1$ and the second as $J_2$.
Then each term of $\E (\xi_{i_1j}\xi_{i_2j'}\xi_{i_3k}\xi_{i_4k'})$ in the expanded summation $\E\{J_1(j,k)-J_2(j,k)\}\{J_1(j',k')-J_2(j',k')\}$ is equal to $0$ as long as $j\not=j'$ or $k\not=k'$,  thus the off-diagonal elements of $\Gamma$ are all zero.
Denote 
\begin{equation}\label{covariance notation}
U_r = \E(\bm\xi_r\bm\xi_r^{\trans}), \, U_{r,1} = \E(\bm\xi_{r,1}\bm\xi^{\trans}_{r,1}), \, U_{r,2} = \E(\bm\xi_{r,2}\bm\xi^{\trans}_{r,2}), \, V_{r,1}= \E(\bm\xi_r\bm\xi_{r,1}^{\trans}), \, V_{r,2}= \E(\bm\xi_r\bm\xi_{r,2}^{\trans})
\end{equation}
We then have $\var(J_1)= {N}^{-1} \E(\bm\xi_j^{\trans}\bm\xi_k \bm\xi_k^{\trans} \bm\xi_j ) = {N}^{-1}\tr(U_jU_k)$, $\var(J_2)=N/(\rho_{jk}^{(h)}N_h)^2~\tr(U_{j,1}U_{k,2})$, $\cov(J_1,J_2)=(\rho_{jk}^{(h)} N_h)^{-1} ~\tr(V_{j,1}V_{k,2}^{\trans}) $,
and it follows that
\begin{align*}
\sigma^2(j,k;j',k') &= \cov\left\{ T_h^*(j,k),T_h^*(j',k') \right\}		\nonumber \\ 
                           &=\delta(j,j') \, \delta(k,k') ~\left\{ \frac{1}{N} \mbox{tr}(U_jU_k) + \frac{N}{(\rho_{jk}^{(h)} N_h)^2} \mbox{tr}(U_{j,1}U_{k,2}) - \frac{2}{\rho_{jk}^{(h)} N_h} \mbox{tr}(V_{j,1}V^{\trans}_{k,2}) \right\}.
\end{align*}
Therefore we have $\Gamma = \mbox{diag} \{\sigma^2_{j, k}: (j, k)\in \mathbb{I}_{R_N}\} $, and let $S'_h( \mathbb{I}_{R_N} ) =\sum_{(j,k) \in  \mathbb{I}_{R_N} } \{T'_h(j,k)/\sigma_{j,k}\}^2.$
Recall that $\| T_h(\mathbb{I}_{R_N}) - T_h'(\mathbb{I}_{R_N}) \| = o_p(1)$ in the proof of Theorem \ref{joint asymptotic normality of Tn}, then $S_h(\mathbb{I}_{R_N}) - S'_h(\mathbb{I}_{R_N}) = o_p(1)$ follows from the continuous mapping theorem applied to the mapping $ \tau_2 \, T_h(\mathbb{I}_{R_N}) = \| \Gamma^{-1/2} T_h(\mathbb{I}_{R_N}) \|^2 $.
As $\pi \left( T'_h(\mathbb{I}_{R_N}), \CZ_{R_N^*} \right) \rightarrow 0$, we then have $\pi \left( \|T'_h(\mathbb{I}_{R_N})\|^2, \| \CZ_{R_N^*} \|^2 \right) \leq \pi(T'_h(\mathbb{I}_{R_N}), \CZ_{R_N^*})[-\log\{\pi(  T'_h(\mathbb{I}_{R_N}), \CZ_{R_N^*} )\} ] ^{1/2} \rightarrow 0$ by the results in Section 4.1 of \cite{mas2002rates}. 
Then it follows that $\pi ( S'_h(\mathbb{I}_{R_N}), \chi^2_{R_N^*} ) \rightarrow 0$ using again the mapping theorem \citep{whitt1974preservation} on $\tau_2$, and thus $\pi ( S_h(\mathbb{I}_{R_N}), \chi^2_{R_N^*} )  \rightarrow 0$.
\end{proof}

   \end{appendix}		

\references
	

@article{fuentes2006testing,
	Author = {Fuentes, Montserrat},
	Journal = {Journal of statistical planning and inference},
	Number = {2},
	Pages = {447--466},
	Publisher = {Elsevier},
	Title = {Testing for separability of spatial-temporal covariance functions},
	Volume = {136},
	Year = {2006}}

@article{mitchell2006likelihood,
	Author = {Mitchell, Matthew W and Genton, Marc G and Gumpertz, Marcia L},
	Journal = {Journal of Multivariate Analysis},
	Number = {5},
	Pages = {1025--1043},
	Publisher = {Academic Press},
	Title = {A likelihood ratio test for separability of covariances},
	Volume = {97},
	Year = {2006}}

@article{li2007a,
	Author = {Li, Bo and Genton, Marc G. and Sherman, Michael},
	Journal = {Journal of the American Statistical Association},
	Number = {478},
	Pages = {736-744},
	Title = {A nonparametric assessment of properties of space-time covariance functions},
	Volume = {102},
	Year = {2007}}

@article{gneiting2006geostatistical,
	Author = {Gneiting, Tilmann and Genton, Marc G and Guttorp, Peter},
	Journal = {Monographs On Statistics and Applied Probability},
	Pages = {151},
	Publisher = {Chapman \& Hall},
	Title = {Geostatistical space-time models, stationarity, separability, and full symmetry},
	Volume = {107},
	Year = {2006}}

@article{aston2017tests,
	Author = {Aston, John AD and Pigoli, Davide and Tavakoli, Shahin and others},
	Journal = {The Annals of Statistics},
	Number = {4},
	Pages = {1431--1461},
	Publisher = {Institute of Mathematical Statistics},
	Title = {Tests for separability in nonparametric covariance operators of random surfaces},
	Volume = {45},
	Year = {2017}}

@article{constantinou2017testing,
	Author = {Constantinou, Panayiotis and Kokoszka, Piotr and Reimherr, Matthew},
	Journal = {Biometrika},
	Number = {2},
	Pages = {425--437},
	Publisher = {Oxford University Press},
	Title = {Testing separability of space-time functional processes},
	Volume = {104},
	Year = {2017}}

@article{lynch2018test,
	Author = {Lynch, Brian and Chen, Kehui},
	Journal = {Biometrika},
	Number = {4},
	Pages = {815--831},
	Publisher = {Oxford University Press},
	Title = {A test of weak separability for multi-way functional data, with application to brain connectivity studies},
	Volume = {105},
	Year = {2018}}

@book{RamsayJ.O.JamesO.2005Fda/,
	Author = {Ramsay, James and Silverman, Bernard},
	Edition = {2nd},
	Publisher = {Springer},
	Title = {Functional Data Analysis},
	Year = {2005}}

@book{Hsing2015Theoretical,
	Author = {Hsing, Tailen and Eubank, Randall},
	Publisher = {John Wiley \& Sons},
	Title = {Theoretical foundations of functional data analysis, with an introduction to linear operators},
	Year = {2015}}

@article{YaoFang2005FDAf,
	Author = {Yao, Fang and M\"uller, Hans-Georg and Wang, Jane-Ling},
	Issn = {01621459},
	Journal = {Journal of the American Statistical Association},
	Language = {eng},
	Month = {June},
	Number = {470},
	Pages = {577--590},
	Publisher = {American Statistical Association},
	Title = {Functional data analysis for sparse longitudinal data},
	Volume = {100},
	Year = {2005}}

@article{LiYehua2010UCRF,
	Author = {Li, Yehua and Hsing, Tailen},
	Issn = {00905364},
	Journal = {The Annals of Statistics},
	Language = {eng},
	Month = {December},
	Number = {6},
	Pages = {3321--3351},
	Publisher = {Institute of Mathematical Statistics},
	Title = {Uniform convergence rates for nonparametric regression and principal component analysis in functional/longitudinal data},
	Volume = {38},
	Year = {2010}}

@article{hall2006properties,
	Author = {Hall, Peter and Hosseini-Nasab, Mohammad},
	Journal = {Journal of the Royal Statistical Society: Series B (Statistical Methodology)},
	Number = {1},
	Pages = {109--126},
	Publisher = {Wiley Online Library},
	Title = {On properties of functional principal components analysis},
	Volume = {68},
	Year = {2006}}

@article{hall2007methodology,
	Author = {Hall, Peter and Horowitz, Joel L.},
	Journal = {The Annals of Statistics},
	Number = {1},
	Pages = {70--91},
	Publisher = {Institute of Mathematical Statistics},
	Title = {Methodology and convergence rates for functional linear regression},
	Volume = {35},
	Year = {2007}}

@inproceedings{hall2009theory,
  title={Theory for high-order bounds in functional principal components analysis},
  author={Hall, Peter and Hosseini-Nasab, Mohammad},
  booktitle={Mathematical Proceedings of the Cambridge Philosophical Society},
  volume={146},
  number={1},
  pages={225--256},
  year={2009},
  organization={Cambridge University Press}
}

@article{kong2016partially,
	Author = {Kong, Dehan and Xue, Kaijie and Yao, Fang and Zhang, Hao H},
	Journal = {Biometrika},
	Number = {1},
	Pages = {147--159},
	Publisher = {Oxford University Press},
	Title = {Partially functional linear regression in high dimensions},
	Volume = {103},
	Year = {2016}}

@article{hormann2010weakly,
	Author = {H{\"o}rmann, Siegfried and Kokoszka, Piotr and others},
	Journal = {The Annals of Statistics},
	Number = {3},
	Pages = {1845--1884},
	Publisher = {Institute of Mathematical Statistics},
	Title = {Weakly dependent functional data},
	Volume = {38},
	Year = {2010}}

@article{li2007nonparametric,
	Author = {Li, Yehua and Wang, Naisyin and Hong, Meeyoung and Turner, Nancy D and Lupton, Joanne R and Carroll, Raymond J},
	Journal = {The Annals of Statistics},
	Number = {4},
	Pages = {1608--1643},
	Title = {Nonparametric estimation of correlation functions in longitudinal and spatial data, with application to colon carcinogenesis experiments},
	Volume = {35},
	Year = {2007}}

@incollection{hormann2011consistency,
	Author = {H{\"o}rmann, Siegfried and Kokoszka, Piotr},
	Booktitle = {Recent Advances in Functional Data Analysis and Related Topics},
	Pages = {169--175},
	Publisher = {Springer},
	Title = {Consistency of the mean and the principal components of spatially distributed functional data},
	Year = {2011}}

@article{gromenko2012estimation,
	Author = {Gromenko, Oleksandr and Kokoszka, Piotr and Zhu, Lie and Sojka, Jan},
	Journal = {The Annals of Applied Statistics},
	Number = {2},
	Pages = {669--696},
	Publisher = {JSTOR},
	Title = {Estimation and testing for spatially indexed curves with application to ionospheric and magnetic field trends},
	Volume = {6},
	Year = {2012}}

@article{liu2017functional,
	Author = {Liu, Chong and Ray, Surajit and Hooker, Giles},
	Journal = {Statistics and Computing},
	Number = {6},
	Pages = {1639--1654},
	Publisher = {Springer},
	Title = {Functional principal component analysis of spatially correlated data},
	Volume = {27},
	Year = {2017}}

@article{paul2011principal,
	Author = {Paul, Debashis and Peng, Jie},
	Journal = {Electronic Journal of Statistics},
	Pages = {1960--2003},
	Publisher = {The Institute of Mathematical Statistics and the Bernoulli Society},
	Title = {Principal components analysis for sparsely observed correlated functional data using a kernel smoothing approach},
	Volume = {5},
	Year = {2011}}

@article{li2014functional,
	Author = {Li, Yehua and Guan, Yongtao},
	Journal = {Journal of the American Statistical Association},
	Number = {507},
	Pages = {1205--1215},
	Publisher = {Taylor \& Francis},
	Title = {Functional principal component analysis of spatio-temporal point processes with applications in disease surveillance},
	Volume = {109},
	Year = {2014}}

@article{ZhouLan2010Rrme,
	Author = {Zhou, Lan and Huang, Jianhua Z and Martinez, Josue G and Maity, Arnab and Baladandayuthapani, Veerabhadran and Carroll, Raymond J},
	Journal = {Journal of the American Statistical Association},
	Number = {489},
	Pages = {390--400},
	Publisher = {Taylor \& Francis},
	Title = {Reduced rank mixed effects models for spatially correlated hierarchical functional data},
	Volume = {105},
	Year = {2010}}

@article{rosenblatt1956central,
	Author = {Rosenblatt, Murray},
	Journal = {Proceedings of the National Academy of Sciences},
	Number = {1},
	Pages = {43--47},
	Publisher = {National Acad Sciences},
	Title = {A central limit theorem and a strong mixing condition},
	Volume = {42},
	Year = {1956}}

@article{clifford1989assessing,
	Author = {Clifford, Peter and Richardson, Sylvia and H{\'e}mon, Denis},
	Journal = {Biometrics},
	Pages = {123--134},
	Publisher = {JSTOR},
	Title = {Assessing the significance of the correlation between two spatial processes},
	Year = {1989}}

@article{dutilleul1993modifying,
	Author = {Dutilleul, Pierre and Clifford, Peter and Richardson, Sylvia and Hemon, Denis},
	Journal = {Biometrics},
	Number = {1},
	Pages = {305--314},
	Publisher = {JSTOR},
	Title = {Modifying the t test for assessing the correlation between two spatial processes},
	Volume = {49},
	Year = {1993}}

@article{gneiting2010matern,
	Author = {Gneiting, Tilmann and Kleiber, William and Schlather, Martin},
	Journal = {Journal of the American Statistical Association},
	Number = {491},
	Pages = {1167--1177},
	Publisher = {Taylor \& Francis},
	Title = {Mat{\'e}rn cross-covariance functions for multivariate random fields},
	Volume = {105},
	Year = {2010}}

@article{guan2004nonparametric,
	Author = {Guan, Yongtao and Sherman, Michael and Calvin, James A},
	Journal = {Journal of the American Statistical Association},
	Number = {467},
	Pages = {810--821},
	Publisher = {Taylor \& Francis},
	Title = {A nonparametric test for spatial isotropy using subsampling},
	Volume = {99},
	Year = {2004}}

@book{sherman2011spatial,
	Author = {Sherman, Michael},
	Publisher = {John Wiley \& Sons},
	Title = {Spatial Statistics and Spatio-temporal Data: Covariance Functions and Directional Properties},
	Year = {2011}}

@book{cressie2015statistics,
	Author = {Cressie, Noel and Wikle, Christopher K},
	Publisher = {John Wiley \& Sons},
	Title = {Statistics for spatio-temporal data},
	Year = {2015}}

@article{li2008testing,
	Author = {Li, Bo and Genton, Marc G and Sherman, Michael},
	Journal = {Biometrika},
	Number = {4},
	Pages = {813--829},
	Publisher = {Oxford University Press},
	Title = {Testing the covariance structure of multivariate random fields},
	Volume = {95},
	Year = {2008}}

@article{LiangXuan2015ABP2,
	Author = {Liang, Xuan and Zou, Tao and Guo, Bin and Li, Shuo and Zhang, Haozhe and Zhang, Shuyi and Huang, Hui and Chen, Song Xi},
	Issn = {1364-5021},
	Journal = {Proceedings of the Royal Society A: Mathematical, Physical and Engineering Science},
	Language = {eng},
	Month = {October},
	Number = {2182},
	Title = {Assessing {B}eijing's {PM}$_{2.5}$ pollution: severity, weather impact, {APEC} and winter heating},
	Volume = {471},
	Year = {2015}}

@article{zhang2017cautionary,
	Author = {Zhang, Shuyi and Guo, Bin and Dong, Anlan and He, Jing and Xu, Ziping and Chen, Song Xi},
	Journal = {Proceedings of the Royal Society A: Mathematical, Physical and Engineering Sciences},
	Number = {2205},
	Pages = {20170457},
	Publisher = {The Royal Society Publishing},
	Title = {Cautionary tales on air-quality improvement in Beijing},
	Volume = {473},
	Year = {2017}}

@article{wu2018probabilistic,
	Author = {Wu, Huangjian and Tang, Xiao and Wang, Zifa and Wu, Lin and Lu, Miaomiao and Wei, Lianfang and Zhu, Jiang},
	Journal = {Advances in Atmospheric Sciences},
	Number = {12},
	Pages = {1522--1532},
	Publisher = {Springer},
	Title = {Probabilistic automatic outlier detection for surface air quality measurements from the China national environmental monitoring network},
	Volume = {35},
	Year = {2018}}

@article{wang2006development,
	Author = {Wang, Zi-fa and Xie, Fu-ying and Wang, Xi-quan and An, J and Zhu, J},
	Journal = {Chinese Journal of Atmospheric Sciences-Chinese Edition},
	Number = {5},
	Pages = {778},
	Publisher = {SCIENCE PRESS},
	Title = {Development and application of nested air quality prediction modeling system},
	Volume = {30},
	Year = {2006}}

@article{Wang2015,
	Author = {Wang, Shuai and Li, Guogang and Gong, Zhengyu and others},
	Day = {01},
	Doi = {10.1007/s11426-015-5468-9},
	Issn = {1869-1870},
	Journal = {Science China Chemistry},
	Month = {Sep},
	Number = {9},
	Pages = {1435--1443},
	Title = {Spatial distribution, seasonal variation and regionalization of {PM}2.5 concentrations in {C}hina},
	Volume = {58},
	Year = {2015}}

@article{zapata2019partial,
	Author = {Zapata, Javier and Oh, Sang-Yun and Petersen, Alexander},
	Journal = {arXiv preprint arXiv:1910.03134},
	Title = {Partial Separability and Functional Graphical Models for Multivariate Gaussian Processes},
	Year = {2019}}

@article{bagchi2020test,
	Author = {Bagchi, Pramita and Dette, Holger and others},
	Journal = {Annals of Statistics},
	Number = {4},
	Pages = {2303--2322},
	Publisher = {Institute of Mathematical Statistics},
	Title = {A test for separability in covariance operators of random surfaces},
	Volume = {48},
	Year = {2020}}

@article{zhang2020unified,
	Author = {Zhang, Haozhe and Li, Yehua},
	Journal = {arXiv preprint arXiv:2006.13489},
	Title = {Unified Principal Component Analysis for Sparse and Dense Functional Data under Spatial Dependency},
	Year = {2020}}

@book{bosq2012nonparametric,
	Author = {Bosq, Denis},
	Publisher = {Springer Science \& Business Media},
	Title = {Nonparametric statistics for stochastic processes: estimation and prediction},
	Volume = {110},
	Year = {2012}}

@book{van2000asymptotic,
	Author = {Van der Vaart, Aad W},
	Publisher = {Cambridge university press},
	Title = {Asymptotic statistics},
	Volume = {3},
	Year = {2000}}

@book{fan1996local,
	Author = {Fan, Jianqing and Gijbels, Irene},
	Publisher = {CRC Press},
	Title = {Local polynomial modelling and its applications: monographs on statistics and applied probability 66},
	Volume = {66},
	Year = {1996}}

@book{billingsley1999convergence,
	Author = {Billingsley, Patrick},
	Publisher = {John Wiley \& Sons},
	Title = {Convergence of probability measures},
	Year = {1999}}

@article{whitt1974preservation,
	Author = {Whitt, Ward},
	Journal = {Zeitschrift f{\"u}r Wahrscheinlichkeitstheorie und Verwandte Gebiete},
	Number = {1},
	Pages = {39--44},
	Publisher = {Springer},
	Title = {Preservation of rates of convergence under mappings},
	Volume = {29},
	Year = {1974}}

@article{mas2002rates,
	Author = {Mas, Andr{\'e}},
	Journal = {Statistics \& probability letters},
	Number = {1},
	Pages = {7--12},
	Publisher = {Elsevier},
	Title = {Rates of weak convergence for images of measures by families of mappings},
	Volume = {56},
	Year = {2002}}
\end{document}